\documentclass{aa}  
\usepackage{graphicx}
\usepackage{txfonts}
\usepackage{color}
\usepackage{natbib}
\usepackage{ulem}
\begin{document}
    \title{Nano-diamonds in proto-planetary discs: 
    }
    \subtitle{Life on the edge}

    \author{A.P. Jones} 

    \institute{Universit\'e Paris-Saclay, CNRS,  Institut d'Astrophysique Spatiale, 91405, Orsay, France.\\
               \email{anthony.jones@universite-paris-saclay.fr}
              }

    \date{Received ? : accepted  ?}


   \abstract
{Nano-diamonds remain an intriguing component of the dust in the few sources where they have been observed in emission.}
{This work focusses on the nano-diamonds observed in circumstellar discs and is an attempt to derive critical information about their possible sizes, compositions, and evolution using a recently-derived set of optical constants.}
{The complex indices of refraction of nano-diamonds and their optical properties (the efficiency factors $Q_{\rm ext}$, $Q_{\rm sca}$, $Q_{\rm abs}$, and $Q_{\rm pr}$) were used to determine their temperatures, lifetimes, and drift velocities as a function of their radii ($0.5-100$\,nm), composition (surface hydrogenation and irradiated states), and distance from the central stars in circumstellar regions.}  
{The nano-diamond temperature profiles were determined for the stars HR\,4049, Elias\,1, and HD\,97048 in the optically-thin limit. The results indicate that large nano-diamonds ($a = 30 - 100$\,nm) are the hottest and therefore the least resistant in the inner disc regions ($\sim 10-50$\, AU), while small ($a < 10$\,nm) fully-hydrogenated nano-diamonds remain significantly cooler in these same regions. We discuss these results within the context of nano-diamond formation in circumstellar discs.}
{Large nano-diamonds, being the hottest, are most affected by the stellar radiation field, however, the effects of radiation pressure appear to be insufficient to move them out of harm's way. The nano-diamonds that best survive and therefore 
shine in the inner regions of proto-planetary discs are then seemingly small ($a < 10$\,nm), hydrogenated, and close in size to pre-solar nano-diamonds ($\langle a \rangle \simeq 1.4$\,nm). Nevertheless, it does not yet appear possible to reconcile their existence there with their seemingly short lifetimes in such regions.}
   \keywords{ISM:abundances -- ISM:dust,extinction               }

    \maketitle
%

\section{Introduction}

The presence of nano-diamonds in the proto-planetary discs around the Herbig Ae/Be stars HD\,97048, and Elias\,1, and in the circumstellar matter around the evolved binary star HR\,4049, is now clearly established through the identification of their characteristic CH and CH$_2$ stretching modes at $3.43$ and $3.53\,\mu$m, respectively, 
which indicate that the nano-diamonds must be relatively large \cite[e.g. $a \simeq 100$\,nm,][]{1999ApJ...521L.133G,2002A&A...384..568V,2004ApJ...614L.129H,2009ApJ...693..610G}. Interestingly, these objects have hot central stars ($T_{\rm eff} \simeq 7,500 - 10,500$\,K), appear to have carbon-rich circumstellar(binary) discs, and the two Herbig Ae/Be stars show x-ray flare activity. These prerequisites appear to be necessary for observable diamond emission and perhaps indicate that there ought to be a source of x-rays in the HR\,4049 stellar system.  Although these sources are relatively well studied observationally the exact location and the origin of the nano-diamonds remains uncertain. 

It is evident that a detailed analysis and interpretation of the properties of these nano-diamonds requires a viable model for their optical properties, one that ideally has some predictive capability. The useful diagnostic potential of any interstellar and circumstellar dust species relies upon an approach informed by laboratory data on analogues as close as possible in composition and structure to those that we infer to exist in space. The usefulness of this laboratory data in turn depends upon the injection of this data into a viable dust modelling framework. A key part of this process, perhaps the key, is the use of the characteristic infrared spectra for interstellar dust analogue materials. In order to fully utilise this potential we need to understand how particle size, composition, and morphology determine the spectral characteristics. In the case of nano-diamonds there exists extensive experimental and modelling data to guide us \citep[e.g.][]{1989Natur.339..117L,1994A&A...284..583C,1995MNRAS.277..986K,1995ApJ...454L.157M,Reich:2011gv,1998A&A...330.1080A,2000M&PS...35...75B,1998A&A...336L..41H,2002JChPh.116.1211C,2002ApJ...581L..55S,2004A&A...423..983M,Jones:2004fu,2007ApJ...661..919P,2011ApJ...729...91S,Usoltseva:2018er,1997PhRvB..55.1838Z}. 

The primary aim of any astronomical nano-diamond study is, at the very least, to be able to estimate their sizes in the objects where they are observed. Seemingly, the best size diagnostic is the ratio of the $3.43\,\mu$m CH and $3.53\,\mu$m CH$_2$ stretching band strengths \citep[e.g.][]{2002JChPh.116.1211C,2002ApJ...581L..55S,2007ApJ...661..919P,2020_Jones_nd_CHn_ratios}. The temperature, size, and morphology dependence of this band ratio has clearly been demonstrated \citep[e.g.][]{2002JChPh.116.1211C,2007ApJ...661..919P,2020_Jones_nd_CHn_ratios} and likely also depends on the nature and degree of the surface hydrogen coverage. Further, and given that nano-diamonds are observed in emission close to hot stars, this ratio will be temperature-dependent \citep[e.g.][]{2002JChPh.116.1211C} because of the underlying thermal emission continuum and  differential de-hydrogenation rates from CH and CH$_2$  surface sites \citep[e.g. see section 8 of][and references therein]{2020_Jones_nd_CHn_ratios}. A lack of sufficiently relevant laboratory data and a wish to keep the modelling as straight-forward as possible means that we will not consider this latter effect.  In order to model, analyse, and interpret the spectra of the nano-diamonds observed in circumstellar regions we make use of the recently-determined nano-diamond structural properties and size-dependent optical constants \citep{2020_Jones_nd_CHn_ratios,2020_Jones_nd_ns_and_ks}. 

Independent of any formation scenario we investigated the survivability of nano-diamonds as a function of their radius, composition, and distance from a star, including the thermal effects that determine their processing and lifetimes, as well as the role of drift velocity in determining their migration within circumstellar regions. We found that at distances of the order of $10-50$\,{\tiny AU} from hot stars ($T_{\rm eff} \sim 7,000 - 10,000$\,K) that $0.5-100$\,nm radius nano-diamonds cannot survive for long but can be stable at larger distances ($R > 15-100$\,{\tiny AU}, depending on the system). Within the context of these results we discussed their possible origin and formation routes. 

This work and its conclusions should, however, be regarded as preliminary given the many remaining uncertainties and unknowns associated with nano-diamonds. Nevertheless, the results probably represent good order of magnitude estimates for the relevant processes and timescales that determine the existence and evolution of nano-diamonds in circumstellar regions. In particular, they reveal discrepancies spanning orders of magnitude, which are unlikely to go away with more sophisticated modelling, and uncover a  circumstellar nano-diamond paradox: The regions where some nano-diamond emission is observed are, apparently, regions where they have extremely short lifetimes. 

The paper is structured as follows: 
Section \ref{sect_properties} is a brief look at the key [CH]/[CH$_2$] ratio\footnote{Square brackets are used in the conventional sense to indicate concentrations or abundances.} in nano-diamond, 
Section \ref{sect_nknanod_model} considers their optical constants, 
Section \ref{sect_optical_props} presents the nano-diamond optical properties,  
Section \ref{sect_Ts} the resulting temperature profiles, 
Section \ref{sect_recon} considers the survivability of nano-diamonds close to hot stars,
Section \ref{sect_notes} enumerates the important points to arise from this work, 
Section \ref{sect_results} discusses the results and speculates upon some of their consequences, 
Section \ref{sect_formation} explores a formation scenario, and 
Section \ref{sect_conclusions} presents the conclusions.

\section{The [CH]/[CH$_2$] ratio in nano-diamonds} 
\label{sect_properties}

In the absence of any dehydrogenation effects, (dis)proportionate or otherwise, \cite{2020_Jones_nd_CHn_ratios} studied the structures and  surface abundances of CH and CH$_2$ groups on the fully-hydrogenated surfaces of (semi-)regular euhedral nano-diamond particles\footnote{The studied euhedral particle shapes were: (semi-)regular tetrahedra, truncated tetrahedra, octahedra, truncated octahedra, cuboctahedra, cubes, and truncated cubes.} with well-defined \{111\} and \{100\} crystalline facets. This study also encompassed spherical nano-diamonds and derived the surface CH$_n$ abundance ratios, [CH]/[CH$_2$], for all particle types as a function of size. This work showed that euhedral nano-diamonds exhibit orders of magnitude dispersion in their [CH]/[CH$_2$] ratios, which increase with size. There is also a large, but more limited, size-to-size dispersion in this ratio ($\sim 0.7-2.1$) for small spherical nano-diamonds ($a < 2$\,nm), which converges to a limit of $\sim 2.4$ for the larger sizes ($a > 2$\,nm). We note that all of the values for spherical nano-diamonds fall well within the measured range ($\sim 0.5-3.0$) for laboratory experiments and astronomical observations. \cite{2020_Jones_nd_CHn_ratios} therefore concluded that spherical particles appear to be the best match to the data and that, further, the best hope of determining nano-diamond sizes from observations is by assuming spherical nano-diamonds and adopting the statistically-averaged, analytical expression for the size-dependence of the [CH]/[CH$_2$] ratio, that is 
\begin{equation}
\frac{{\rm [CH]}}{{\rm [CH_2]}} = 2.265 \, \left( \frac{ a_{\rm nd}}{ 1\,{\rm nm} }\right)^{0.03} - \frac{ 1 }{ 2.5 } \left( \frac{ a_{\rm nd}}{ 1\,{\rm nm} } \right)^{-1}. 
\label{eq_ratio_fit}
\end{equation}
This expression gives a reasonable fit to the inevitable ups and downs of the diamond network calculations at small sizes ($a < 5$\,nm). These calculations were based upon the typical C$-$C bond length in diamond (0.154\,nm) and so are restricted to assuming the $3.52$\,g\,cm$^{-3}$ bulk density for diamond \cite{2020_Jones_nd_CHn_ratios}.

\section{Nano-diamond optical constants}
\label{sect_nknanod_model}

The optical properties of nano-diamonds were measured (absorbance spectra or mass absorption coefficients) in numerous studies \citep{1998A&A...330.1080A,2000M&PS...35...75B,1994A&A...284..583C,1998A&A...336L..41H,Jones:2004fu,1995MNRAS.277..986K,1995ApJ...454L.157M,Reich:2011gv,2002ApJ...581L..55S,2011ApJ...729...91S,Usoltseva:2018er,1997PhRvB..55.1838Z}. Interstellar and circumstellar dust modelling requires well-determined properties (i.e. the complex indices of refraction) over a very wide wavelength range  (i.e. from EUV to mm). More specifically, the optical constants of the pre-solar nano-diamonds extracted from meteorites were measured over wide wavelength ranges  \citep[$\lambda = 0.1 - 1\,\mu$m,][]{1989Natur.339..117L} and \citep[$\lambda = 0.12 - 100 \,\mu$m,][]{2004A&A...423..983M}. Unfortunately, these data do not cover a sufficiently-wide wavelength range to permit the detailed modelling of nano-diamonds in interstellar and circumstellar media but they are eminently suitable for calibrating the models. 

As an aid to astrophysical dust modelling and the interpretation of nano-diamond observations their structural properties were recently investigated by \cite{2020_Jones_nd_CHn_ratios}. Their optical constants, the complex indices of refraction $m(n,k)$, were determined as a function of wavelength, $\lambda$, radius, $a$, fractional surface hydrogenation, $f_{\rm H}$, and bulk diamond structure (pristine or defected) by \cite{2020_Jones_nd_ns_and_ks}. This derivation of the nano-diamond $n$ and $k$ values was based on the optEC$_{\rm (s)}$(a) methodology \citep{2012A&A...540A...1J,2012A&A...540A...2J,2012A&A...542A..98J}, which was developed for a wide compositional range of hydrocarbonaceous materials, a-C(:H). The hydrogenated amorphous carbons encompassed by the optEC$_{\rm (s)}$(a) model includes the sp$^3$-rich, wide band gap ($E_{\rm g} \simeq 2.7$\,eV), hydrogenated amorphous carbons (a-C:H), which can be considered as the closest approach to nano-diamonds within that framework. 

In the determination of nano-diamond optical constants from extreme-UV to mm wavelengths \citep{2020_Jones_nd_ns_and_ks} exact solutions are obtainable for the particle structures by assuming a perfect diamond lattice \citep{2020_Jones_nd_CHn_ratios}, rather than having to adopt something akin to the statistical extended Random Covalent Network (eRCN) and Defective Graphite (DG) descriptions that were developed for a-C(:H) solids. Nevertheless, as mentioned above, at some level we are forced to assume a statistical averaging over the many close-in-size nano-diamond forms possible at small sizes \citep[$a \lesssim 5$\,nm,][]{2020_Jones_nd_CHn_ratios}, a covalent nano-diamond size-randomisation approximation, which is tangentially equivalent to the eRCN and DG approaches. 

The application of the methods and  approximations, described in detail in \cite{2020_Jones_nd_ns_and_ks}, allowed for a determination of the nano-diamond optical constants over more than seven  orders of magnitude in energy, $2.5 \times 10^{-6}$ -- $56$\,eV (i.e. equivalent to 50\,cm -- $0.022\,\mu$m in wavelength, respectively) and over a parameter space of particular relevance to the interpretation of nano-diamond observations. As with all theoretical modelling constructs these data should be considered as adaptable and fine-tunable as and when more-constraining laboratory data and astrophysical observations on (nano-)diamonds become available. As per the optEC$_{\rm (s)}$(a) model, the methodology adopted by \cite{2020_Jones_nd_ns_and_ks} has its foundations firmly planted in the laboratory-measured properties of diamond available at the time. 

The primary aim of these new nano-diamond optical constants is to provide a set of self-consistent data that can be used to qualitatively, and hopefully quantitatively, test the size- and composition-dependent effects of nano-diamond evolution within the astrophysical context. To this end ASCII files of the $n$ and $k$ data for nano-diamond particles are available\footnote{From the following website: 
https://www.ias.u-psud.fr/themis/} for a limited set of particle radii (0.5, 1, 3, 10, 30, and 100\,nm $\equiv 80$, 600, $2 \times 10^4$, $7 \times 10^5$, $2 \times 10^7$, $7 \times 10^8$ C atoms per particle, respectively), which covers particles from the large molecule domain to bulk materials: six $n$ and $k$ data files are provided for three different hydrogenation states ($f_{\rm H} = 0, 0.25$ and 1) and, in each case, for non-irradiated and irradiated nano-diamond particle cores.

\begin{figure}
\includegraphics[width=9.0cm]{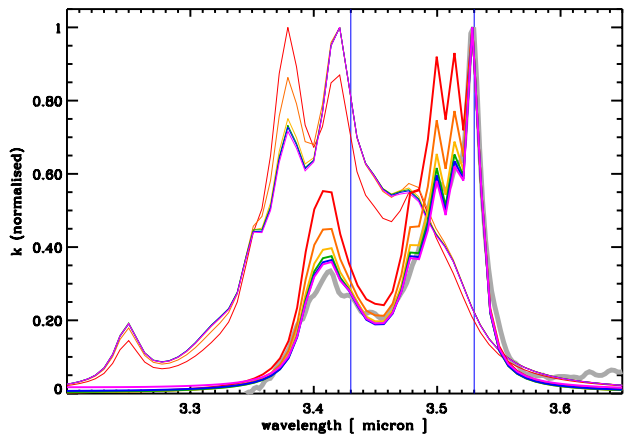}
\caption{The imaginary part of the complex refractive index, $k$, for nano-diamond (thick) and a-C:H (thin) particles,  normalised to their peak values, as a function of radius (0.5, 1, 3, 10, 30, and 100\,nm: red, orange, yellow, green, blue, and violet lines, respectively). The thick grey line shows the 100\,nm diamond data from \cite{Jones:2004fu}.}
\label{fig_spect_nanod2}
\end{figure}

\section{Nano-diamond optical properties}
\label{sect_optical_props}

The nano-diamond optical properties, plotted as $Q_{\rm ext}/a$ and $Q_{\rm abs}/a$ in units of nm$^{-1}$, for particle radii of 0.5, 1, 3, 10, 30, and 100\,nm are shown in Fig.~5 of the companion paper \citep{2020_Jones_nd_ns_and_ks}, the interested reader should refer to that figure for the details. Given Kirchhoff's law (emissivity = absorbtivity) it is the absorption efficiency factor, $Q_{\rm abs}$, that primarily determines the nano-diamond temperatures. In this work we used the nano-diamond optical properties, $Q_{\rm abs}$ ($= Q_{\rm em}$), to derive the likely temperatures of nano-diamonds close to hot stars. 

As an indication of the nano-diamond spectral properties Fig.~\ref{fig_spect_nanod2} shows a zoom of the imaginary part of the refractive index, $k$, in the $3.20-3.65\,\mu$m region \citep{2020_Jones_nd_ns_and_ks}. In the figure each spectrum was normalised to its peak value in this wavelength region. For comparison the thick grey line shows the 100\,nm diamond laboratory data taken from \cite{Jones:2004fu} and upon which the model was fine-tuned. The model nano-diamond spectra clearly show the $3.43\,\mu$m  $3.53\,\mu$m bands, and most of the sub-structures, that have been observed in several astronomical objects. Over this wavelength region, and in their normalised form, the band shapes are independent of $f_{\rm H}$ and bulk radiation damage. 

Note that the form of the spectrum in the $3-4\,\mu$m wavelength region, and in particular, the ratio of the intensities of the $3.43\,\mu$m  $3.53\,\mu$m bands is dependent upon the particle size because the CH and CH$_2$ surface abundances for spherically-approximated and  statistically-averaged nano-diamond particles are size-dependent  \citep{2020_Jones_nd_CHn_ratios}. It should also be noted that the model assumes proportionate dehydrogenation from CH and CH$_2$ groups, that is the [CH]/[CH$_2$] ratio is independent of the degree of dehydrogenation, which may not be strictly valid.

\section{Nano-diamond temperatures and stability}
\label{sect_Ts}

\begin{figure*}
\centering
\begin{center} $
\begin{array}{cc}
   \includegraphics[width=9.0cm]{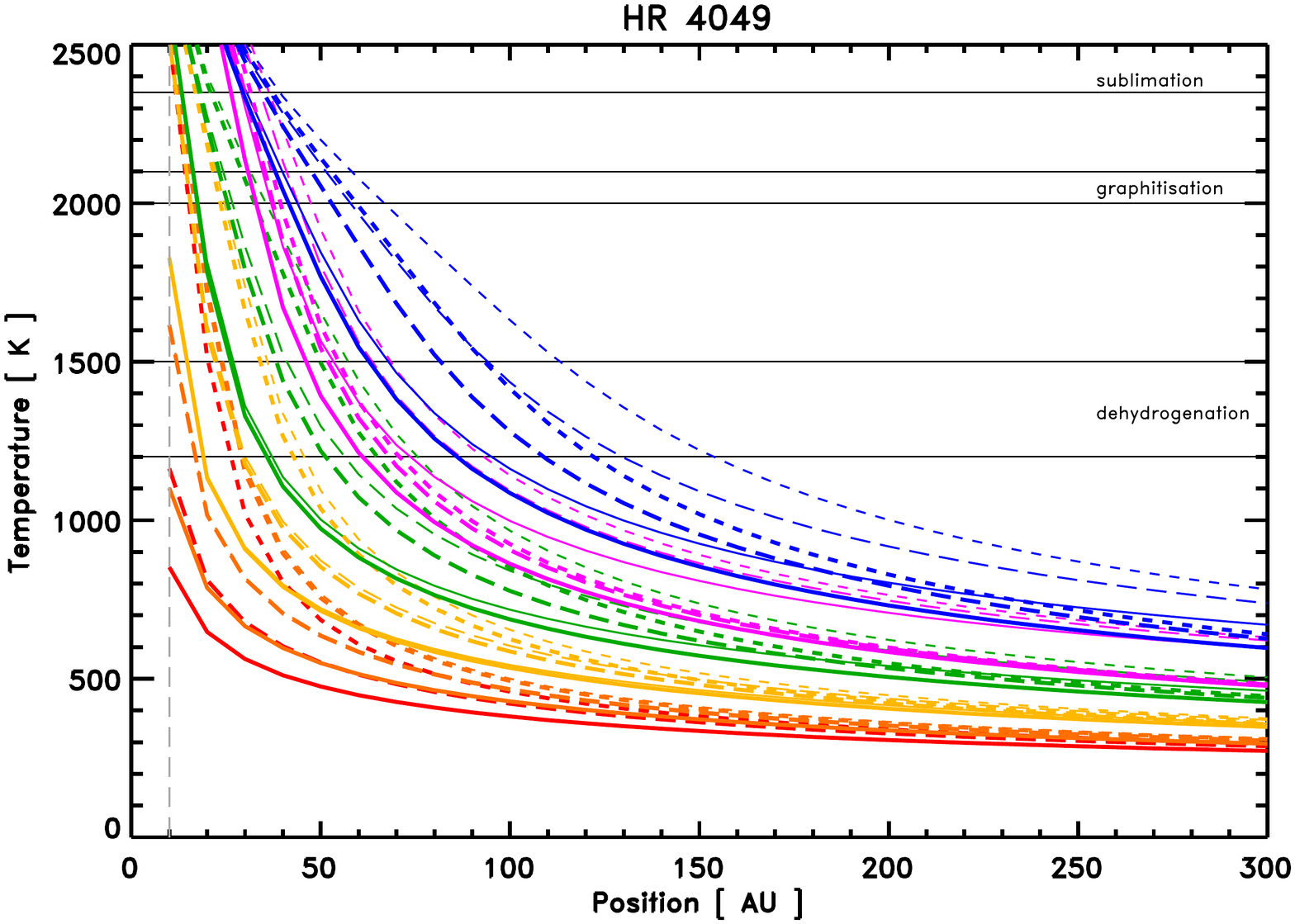}
   \includegraphics[width=9.0cm]{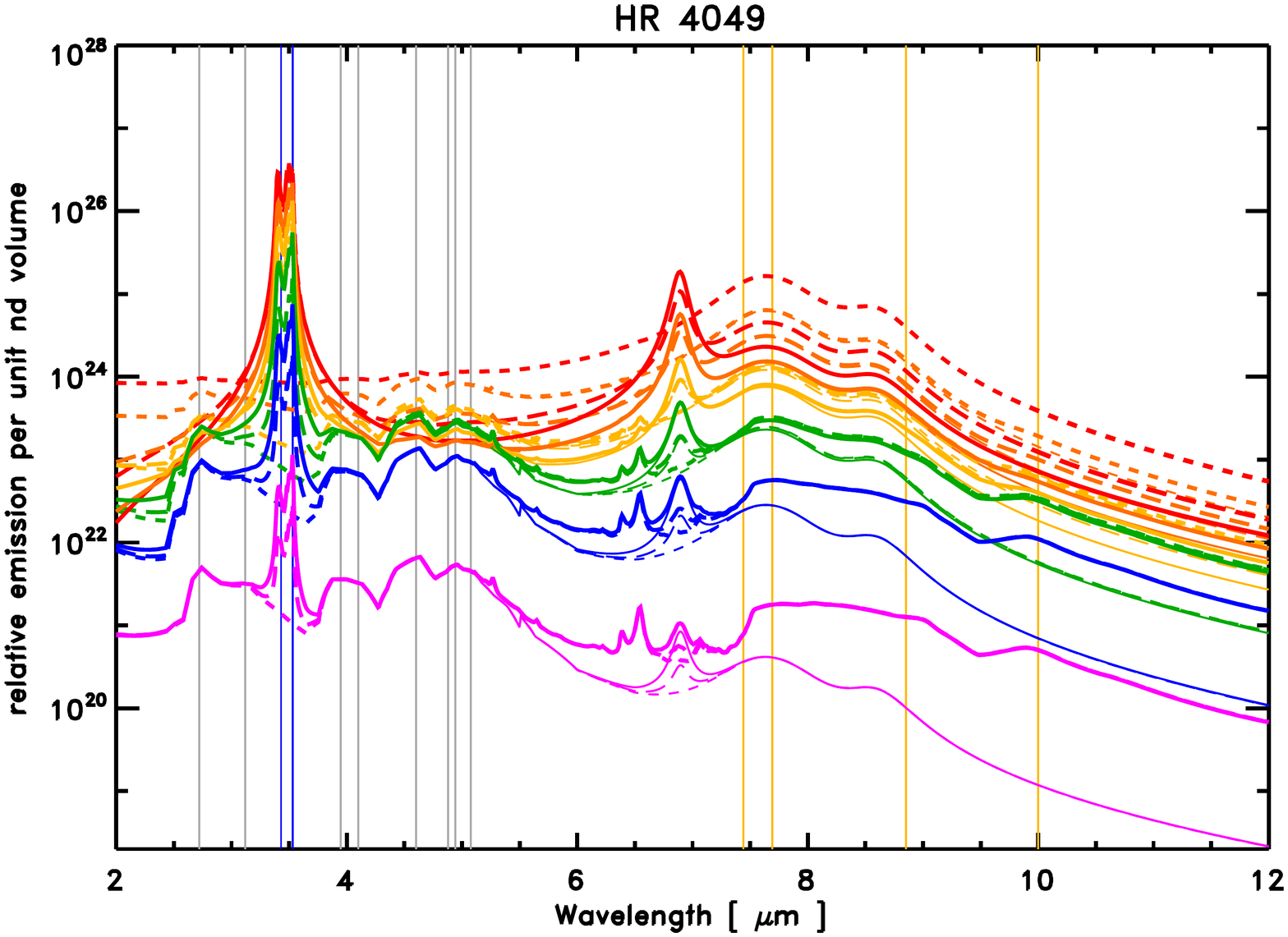}
\end{array} $
\end{center}
\hspace*{-0.9cm}
\begin{center} $
\begin{array}{cc}
   \includegraphics[width=9.0cm]{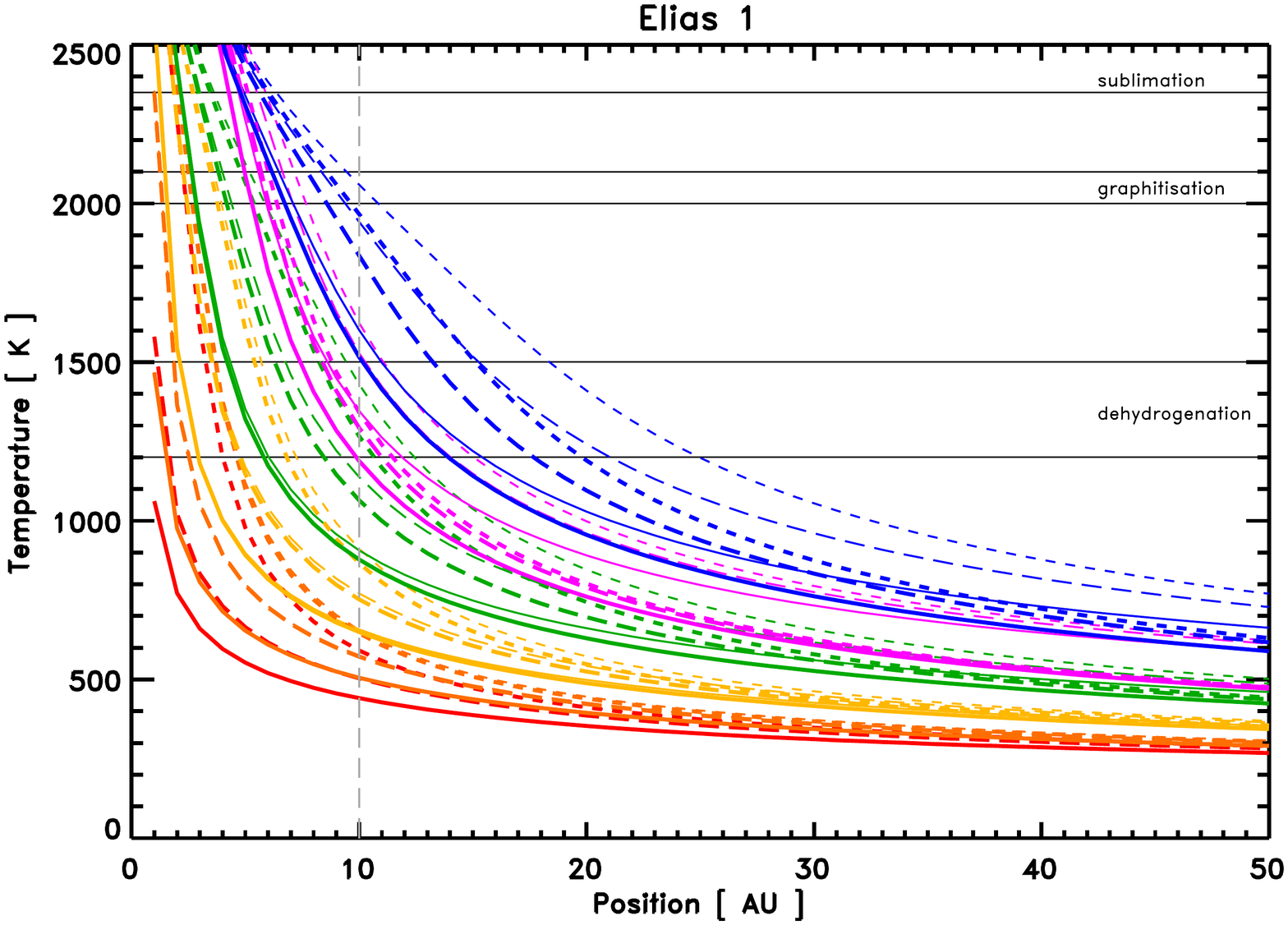}
   \includegraphics[width=9.0cm]{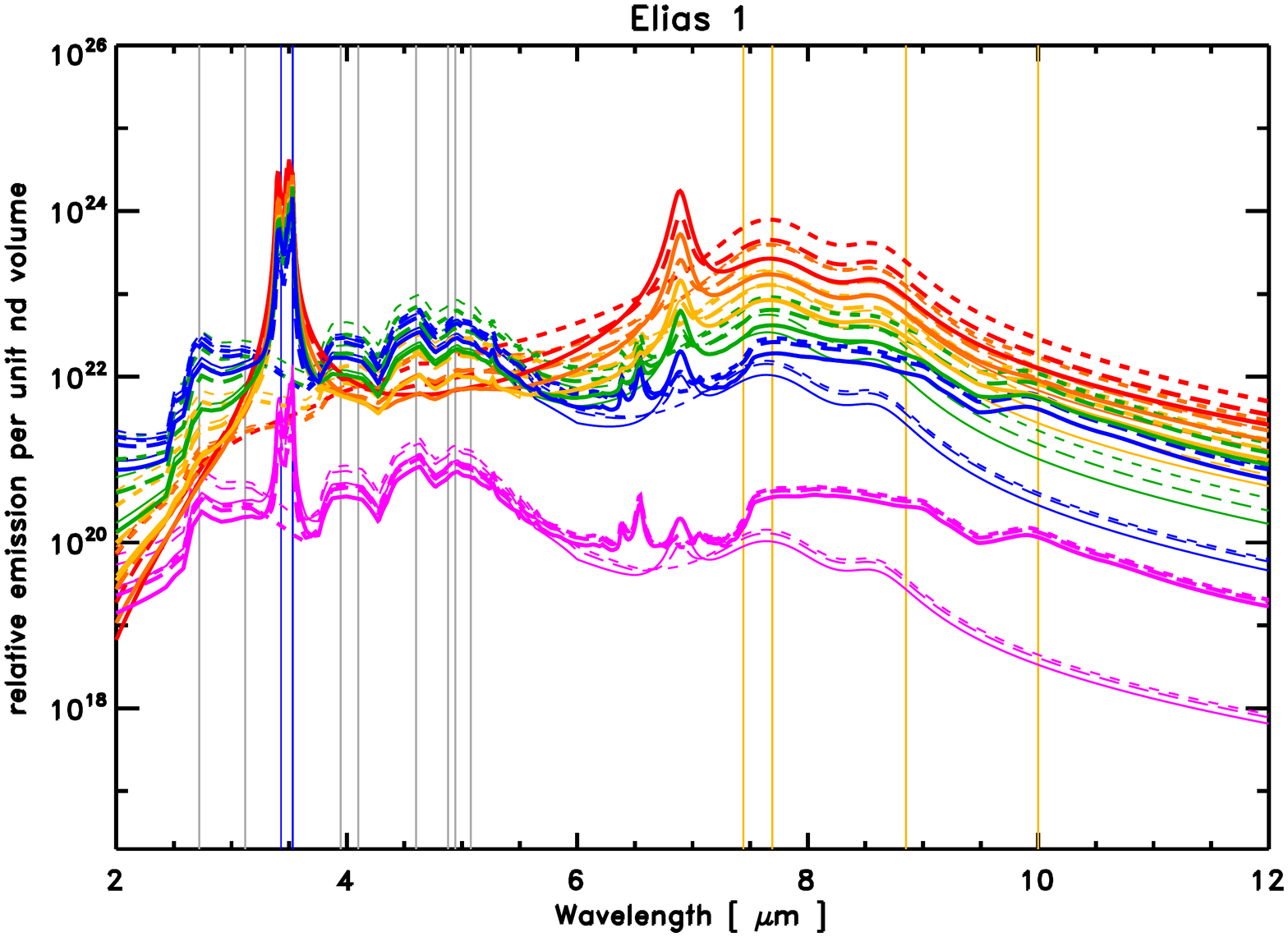}
\end{array} $
\end{center}
\begin{center} $
\begin{array}{cc}
   \includegraphics[width=9.0cm]{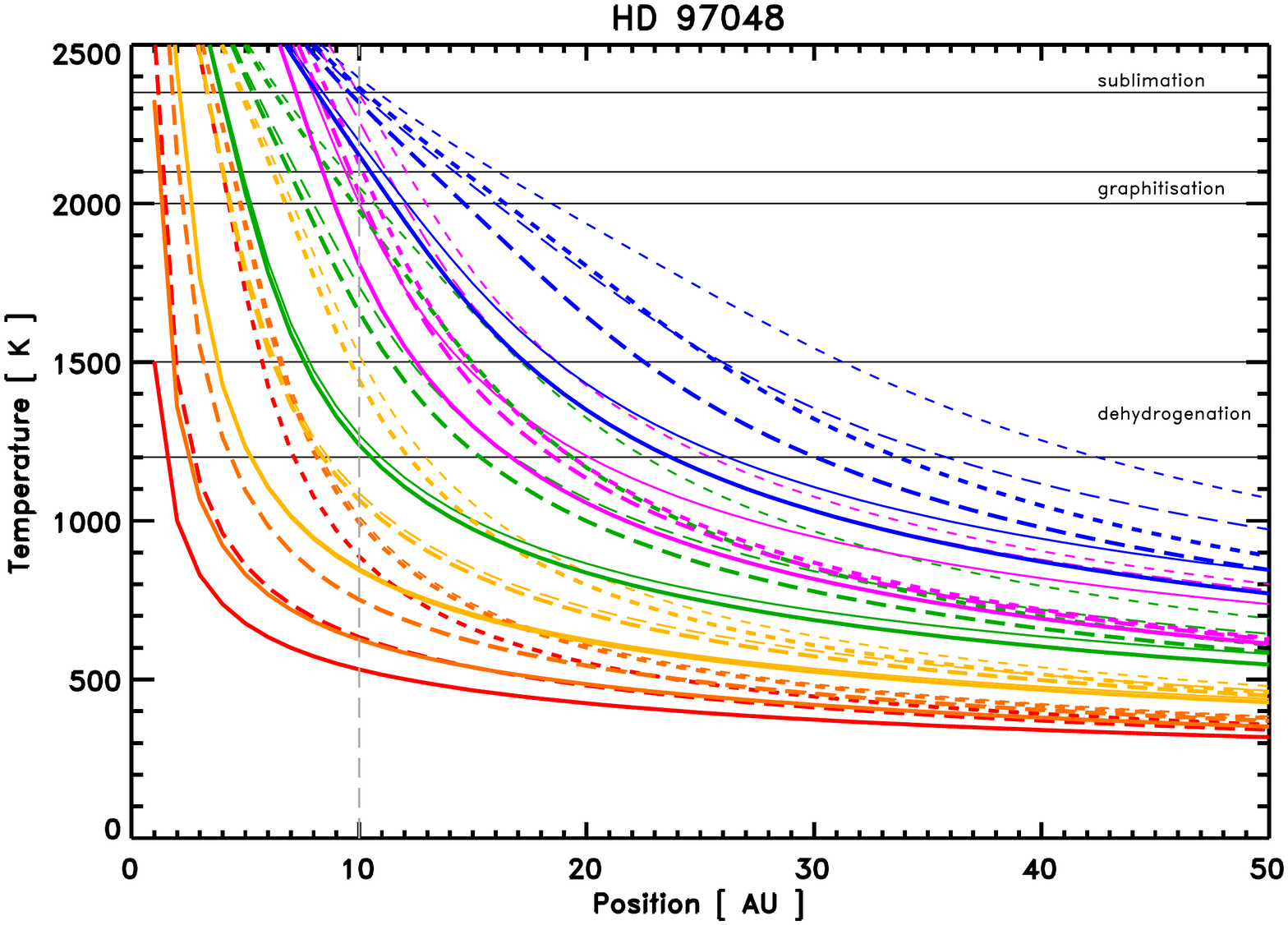} 
   \includegraphics[width=9.0cm]{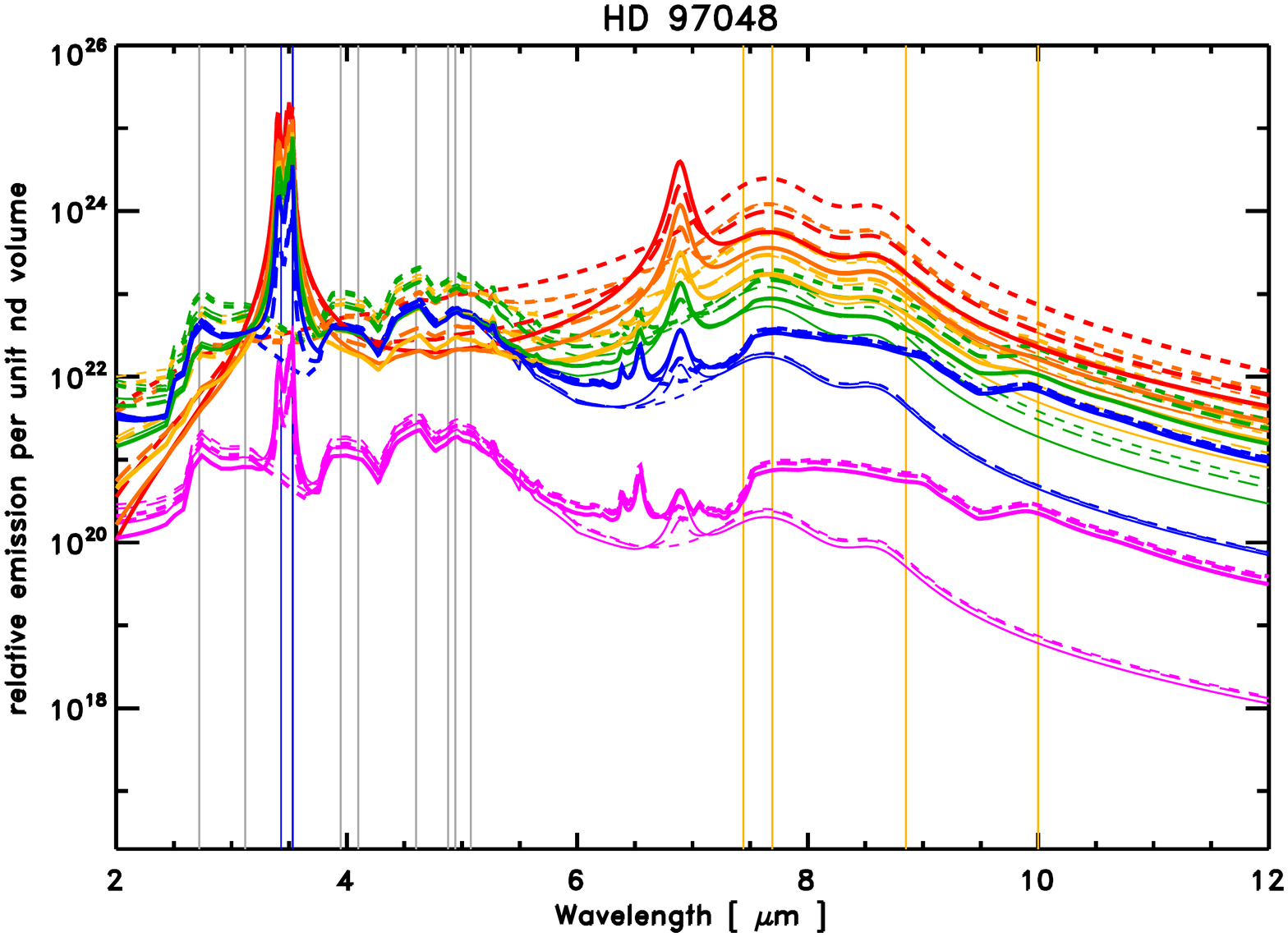}
\end{array} $
\end{center}
\caption{Nano-diamond temperatures for HR\,4049 (top),  Elias\,1 (middle), and HD\,97048 (bottom) as a function of particle radius (0.5, 1, 3, 10, 30, and 100\,nm: red, orange, yellow, green, blue, and violet lines, respectively) and distance from the central star (left panels). {\bf Note the different position axis values for  HD\,97048.} The thinner lines are for non-irradiated diamond and the solid, long-dashed and short-dashed line are for fully-hydrogenated, 75\% dehydrogenated and fully-dehydrogenated nano-diamond surfaces. The right hand panels show the relative nano-diamond emission per unit volume 10\,{\tiny AU} from the star or 50\,{\tiny AU} from the star as in the case of HR\,4049.} 
\label{fig_ndT_profiles}
\centering
\end{figure*}

The left-hand panels in Fig. \ref{fig_ndT_profiles} show the nano-diamond temperatures as a function of particle radius and distance from the central star for the three sources where nano-diamonds have been unequivocally detected (HR\,4049,  Elias\,1, and HD\,97048: top to bottom), for each of the six derived sets of optical properties. The temperatures are calculated in the optically-thin limit, assuming that the nano-diamonds are in thermal equilibrium with the stellar radiation field, that is by solving the usual equation for the balance between the absorbed and emitted energy in a uni-directional radiation field, 
\begin{equation} 
\frac{R_\star^2}{d^2} \int_{0}^\infty Q_{abs}(a_{\rm nd},\lambda) \ \pi B_\lambda(T_{\rm eff}) \ d\lambda 
= 4 \pi \, \langle Q_{abs}(a_{\rm nd},\lambda,T_{\rm nd}) \rangle \, \frac{\sigma}{\pi} \, T_{\rm nd}^4, 
\label{eq_Teq} 
\end{equation}
where $R_\star$ is the stellar radius, $d$ is the distance of the grain from the star, $T_{\rm eff}$ is the stellar effective temperature, $Q_{abs}(a_{\rm nd},\lambda)$ is the nano-diamond absorption efficiency at wavelength $\lambda$, $\langle Q_{abs}(a_{\rm nd},\lambda,T_{\rm nd}) \rangle$ is the Planck-averaged absorption efficiency\footnote{
\[ \langle Q_{abs}(a_{\rm nd},T_{\rm nd}) \rangle =  \frac{ \int_0^\infty Q_{abs}(a_{\rm nd},\lambda) \, B_\lambda(T_{\rm nd}) \, d\lambda }{ \int_0^\infty B_\lambda(T_{\rm nd})  \, d\lambda }  \]
\begin{equation} \hspace*{2.0cm} = \frac{ \pi}{ \sigma \, T_{\rm nd}^4 } \int_0^\infty Q_{abs}(a_{\rm nd},\lambda) \, B_\lambda(T_{\rm nd}) \, d\lambda \end{equation}} and $\sigma$ is the Stefan-Boltzmann constant. The left hand side of Eq. (\ref{eq_Teq}) is the grain heating rate by stellar photon absorption and the right hand side its cooling rate due to thermal emission. The stellar parameters assumed for each source are given in Table \ref{tab_stars}. Note the particularly large radius of the primary in the HR\,4049 system \citep{Acke:2013cn}, which results in a significantly more intense radiation field in this system. 

\begin{figure*}
\centering
\begin{center} $
\begin{array}{cc}
   \includegraphics[width=9.0cm]{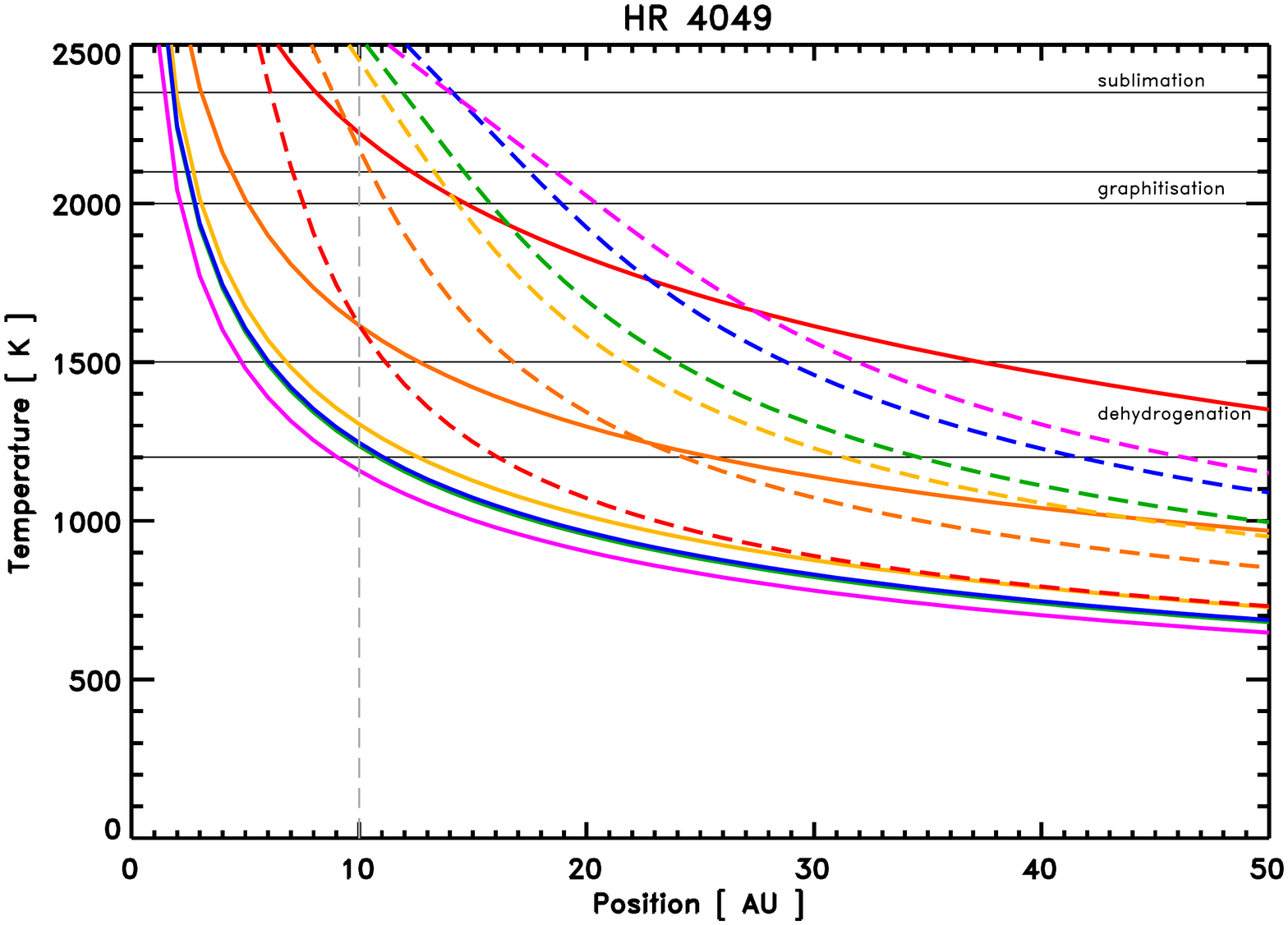}
   \includegraphics[width=9.0cm]{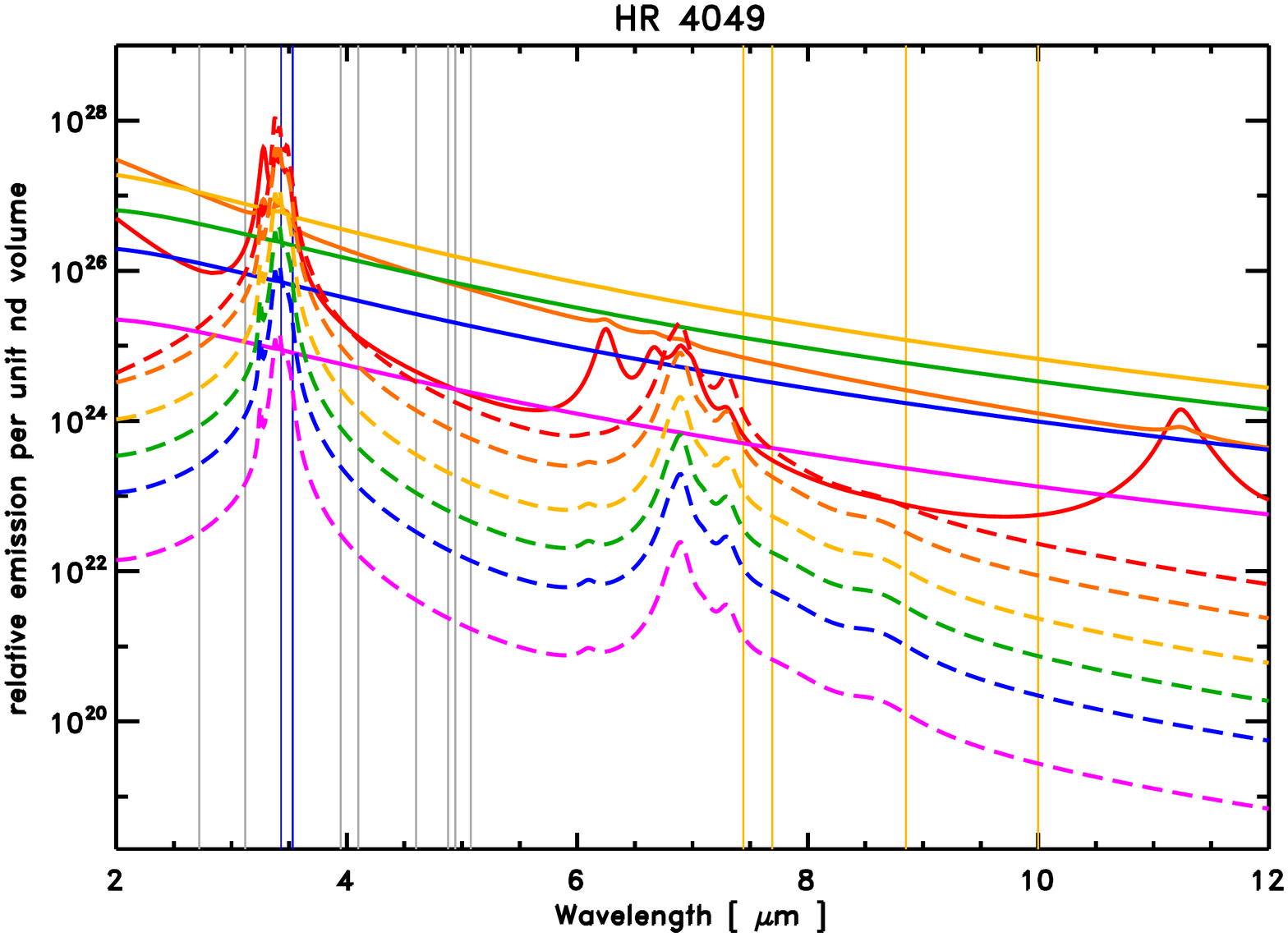}
\end{array} $
\end{center}
\hspace*{-0.9cm}
\begin{center} $
\begin{array}{cc}
   \includegraphics[width=9.0cm]{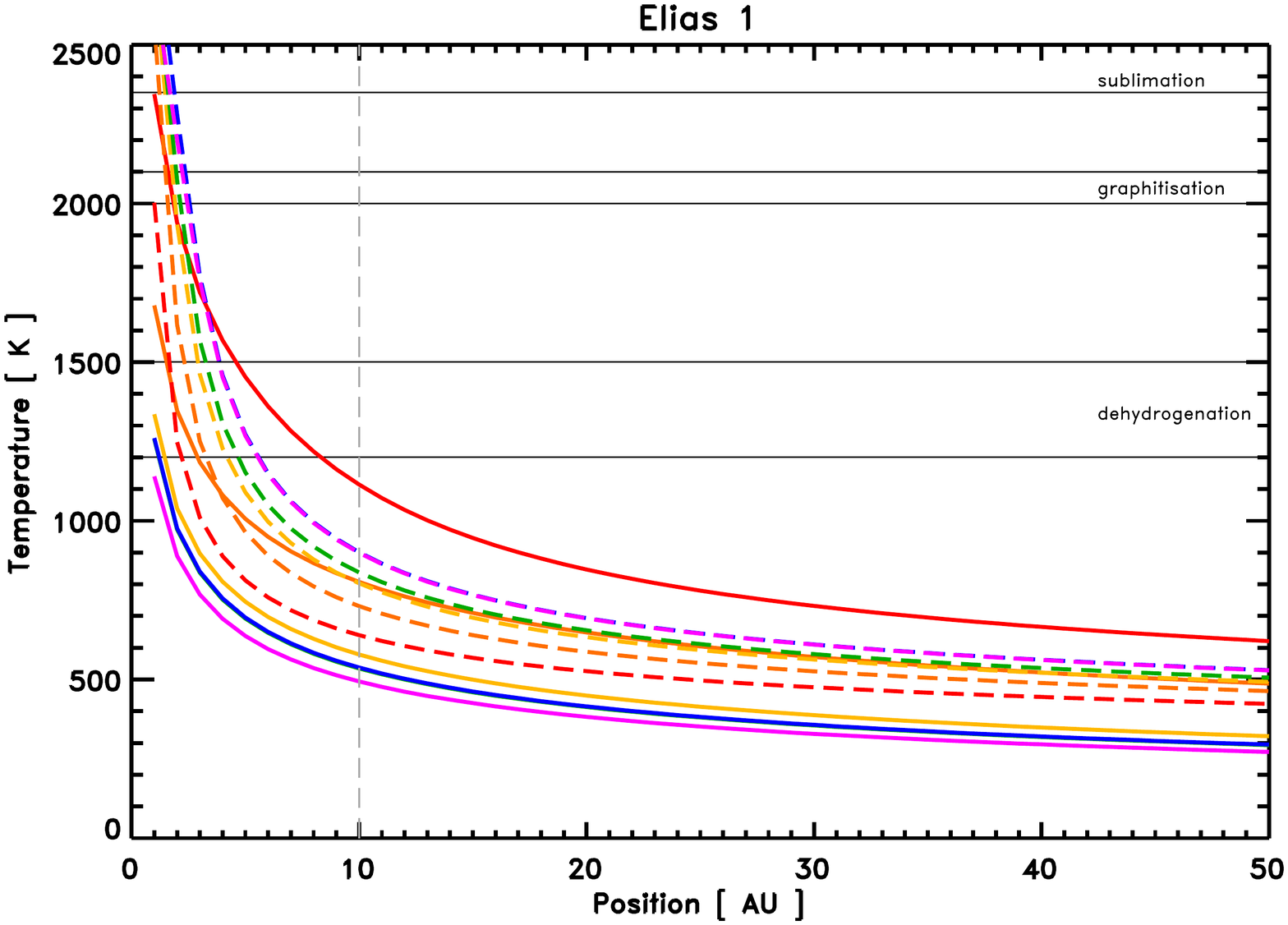}
   \includegraphics[width=9.0cm]{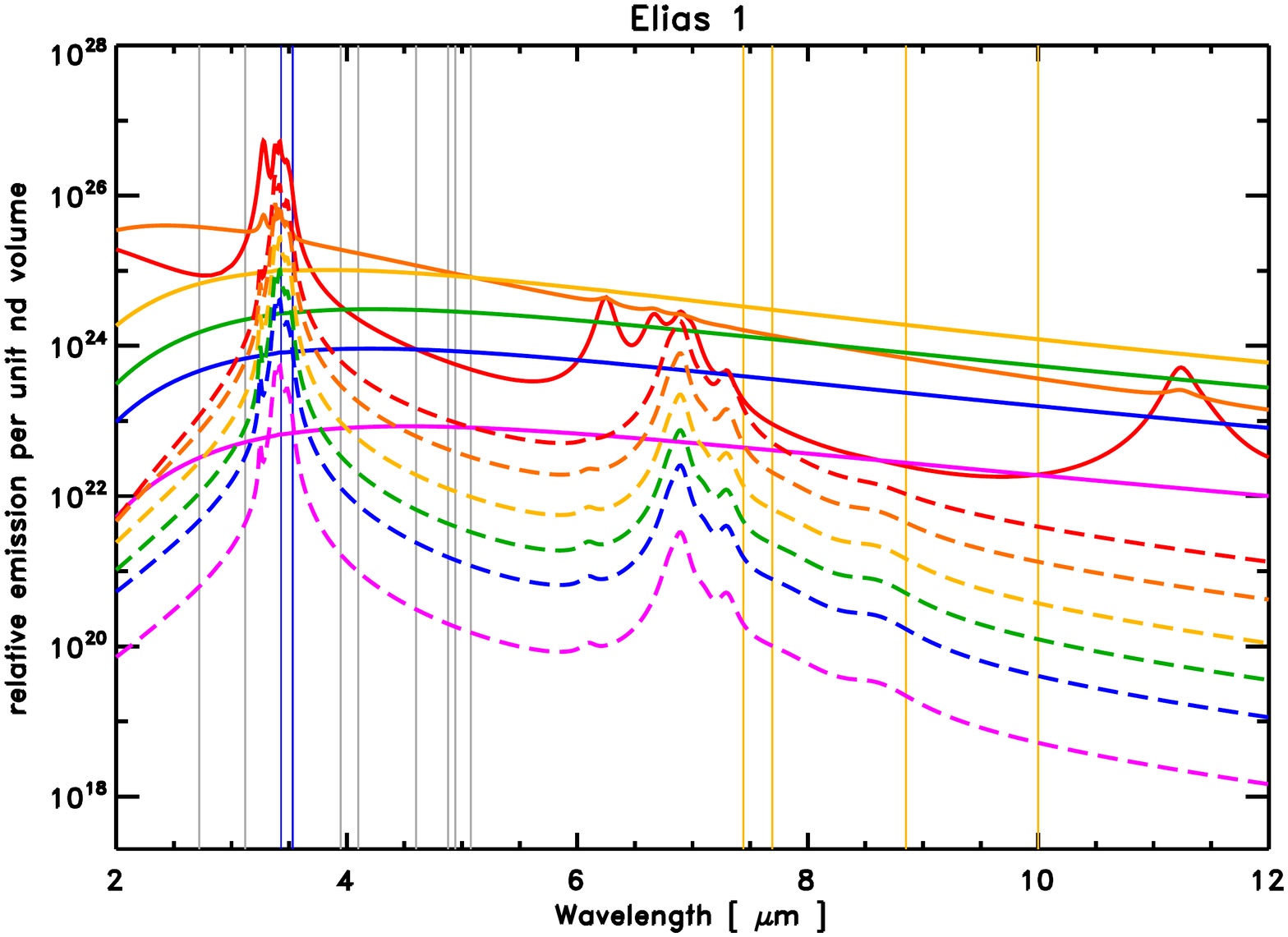}
\end{array} $
\end{center}
\begin{center} $
\begin{array}{cc}
   \includegraphics[width=9.0cm]{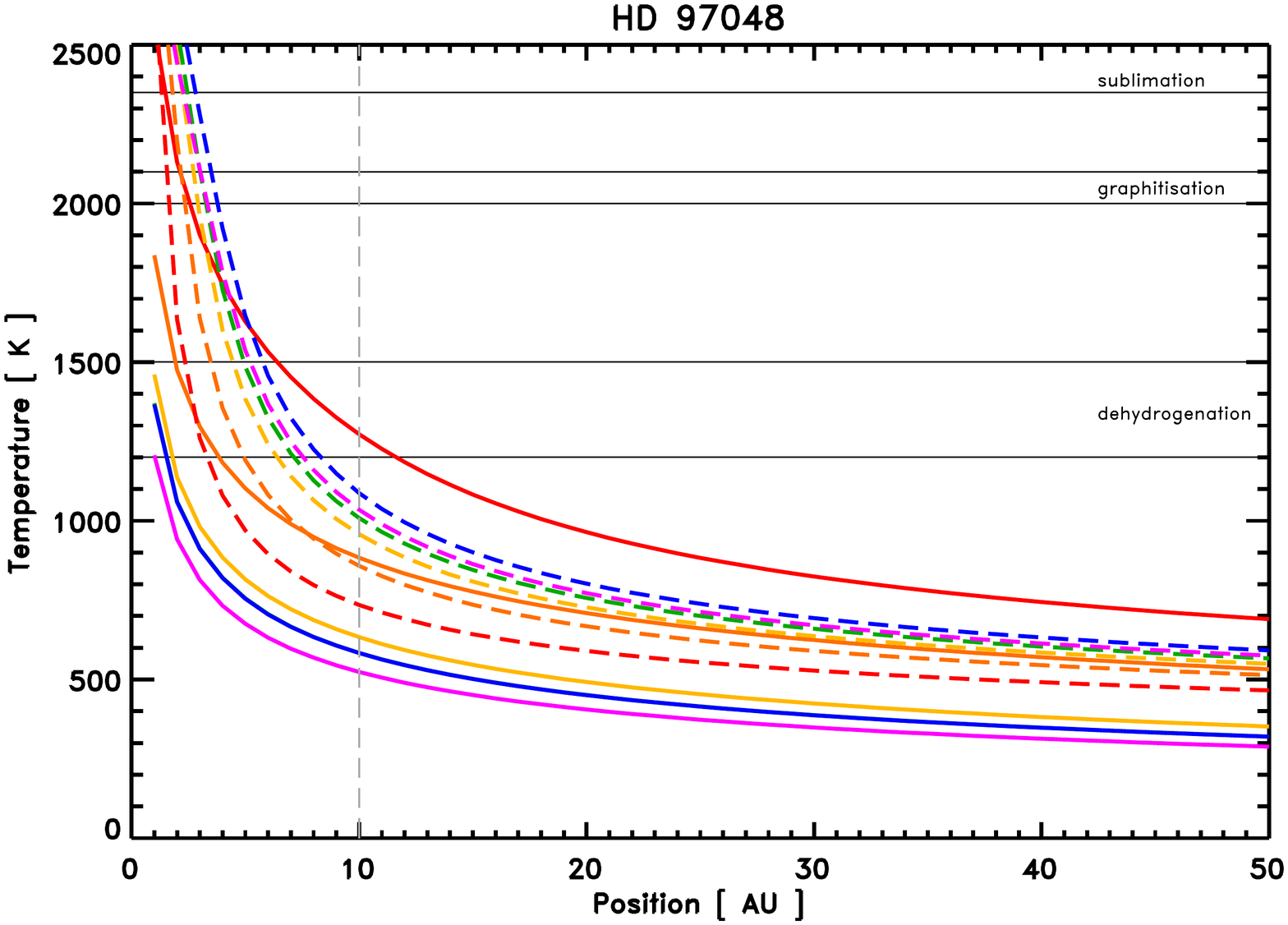} 
   \includegraphics[width=9.0cm]{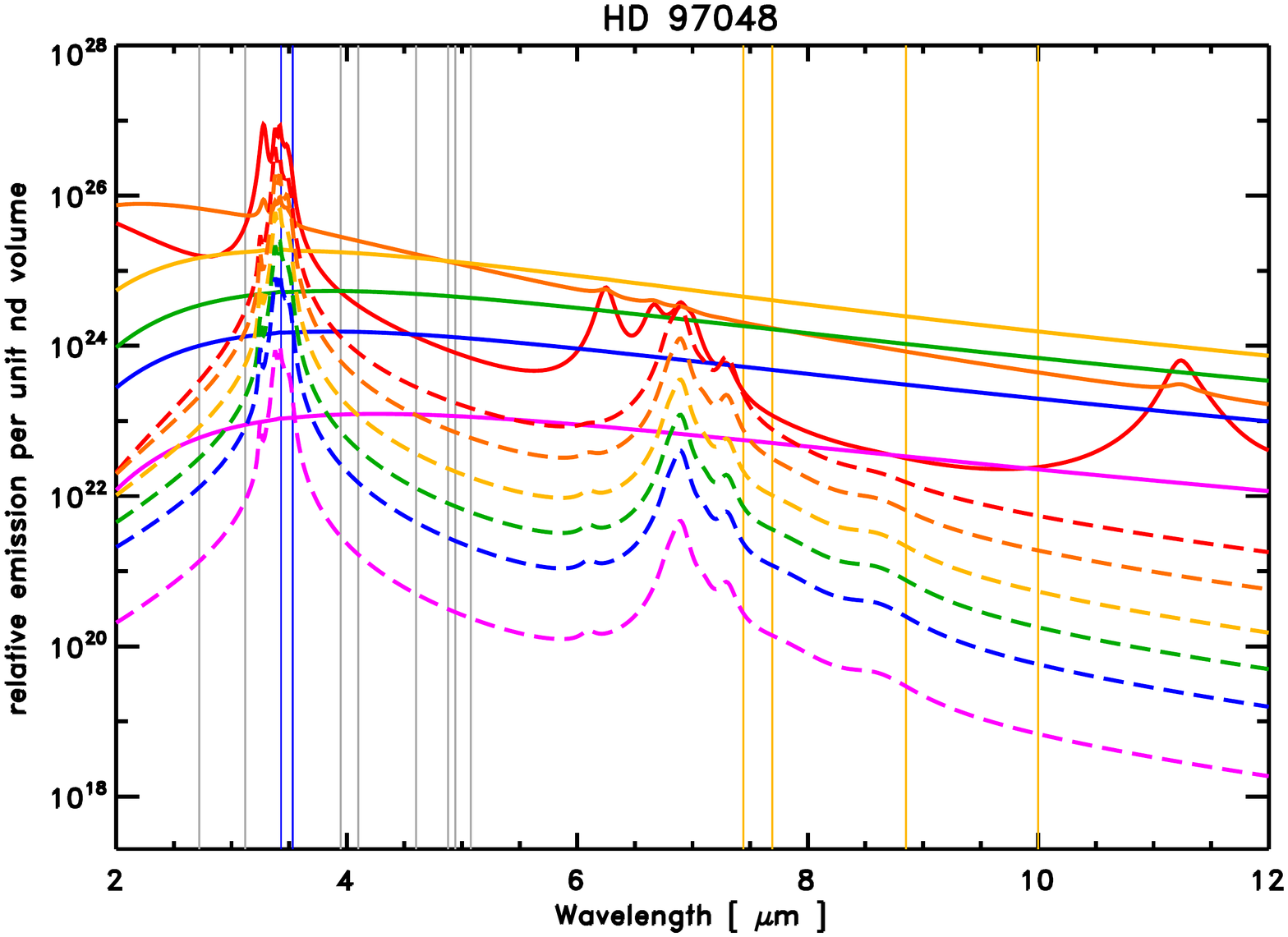}
\end{array} $
\end{center}
\caption{a-C (solid) and a-C:H (dashed) grain temperatures for HR\,4049 (top),  Elias\,1 (middle), and HD\,97048 (bottom) as a function of particle radius (0.5, 1, 3, 10, 30, and 100\,nm: red, orange, yellow, green, blue, and violet lines, respectively) and distance from the central star (left panels). The right hand panels show the relative a-C and a-C:H nano-particle emission per unit volume 10\,{\tiny AU} from the star.}
\label{fig_aCH_profiles}
\centering
\end{figure*}

The nano-diamond temperature profiles for the three sources are shown in Fig.~\ref{fig_ndT_profiles} and for comparison those for aromatic-rich a-C and aliphatic-rich a-C:H grains are given in Fig.~\ref{fig_aCH_profiles}.  The nano-diamond, a-C and a-C:H grain temperatures are given in Table \ref{tab_Tnd}; given its more intense radiation field the HR\,4049 dust temperatures are given at 50\,{\tiny AU}, while those for Elias\,1 and HD\,97048 are given at 10\,{\tiny AU}.

\begin{table}
\caption{Stellar parameters assumed in the modelling.}
\centering                  
\begin{tabular}{l c c c c}        
\hline\hline \\[-0.25 cm]                 
source & $T_{\rm eff}$\,[K] &  $B_\lambda (T_{\rm eff})_{\rm peak}$ [$\mu$m]    & $R_\star/R_\odot$  & $M_\star/M_\odot$  \\    
\hline \\[-0.25 cm]            
HR\,4049$^\star$   &   7,500     &   $\sim$\,0.39   &  31  &  0.4 \\   
Elias\,1      &   9,000     &    $\sim$\,0.32   &  2.0  &  3.5 \\     
HD\,97048 &   10,500   &   $\sim$\,0.28   & 1.7  &  2.5 \\  
\hline                                   
\end{tabular}
\tablefoot{$\star$ Stellar parameters from \cite{Acke:2013cn}.}
\label{tab_stars}
\end{table}

The assumption of thermal equilibrium in the nano-diamond temperature calculations is valid for the larger particles, that is $a_{\rm nd} \gtrsim 10$\,nm. For smaller particles the thermal equilibrium temperatures are close to the average temperatures but the nano-diamond peak temperatures can reach many hundreds of degrees Kelvin.\footnote{Stochastically-heated nano-diamond calculations show that the maximum-to-minimum temperature dispersions, $\delta T$, in the inner disc regions are likely to be less than 200\,K for all of the considered nano-diamonds. This is of the same order, or considerably less, than the dispersion in the thermal equilibrium temperatures induced by the uncertainties in the nano-diamond surface hydrogenation and/or irradiated state (see Fig\,\ref{fig_ndT_profiles}).} The exact calculation of thermal processing timescales in the stochastic heating regime thus requires integrating processes, such as de-hydrogenation, re-construction, and sublimation, over myriad thermal fluctuations extending over many hundreds of degrees Kelvin in matters of only seconds.  Dust sublimation under such thermally fluctuating grain conditions was studied by \cite{1989ApJ...345..230G} for much less intense radiation fields than considered here. Such an approach is beyond the intended scope of the current work and probably premature given the poorly constrained state of the nano-diamond properties in discs. In the following discussions, the estimated smaller nano-diamond ($a_{\rm nd} \lesssim 10$\,nm) lifetimes against thermally-driven processes may have therefore been over- or under-estimated because the thermal fluctuations extend both below and above the calculated equilibrium temperatures. Thus, the derived lifetimes will, in reality, depend upon the fine details of the temperature fluctuations. In the following we will, nevertheless and where relevant, discuss the limitations and implications of the thermal equilibrium assumption. Further, and given that the observed nano-diamond emission bands appear to be consistent with relatively large nano-diamonds \cite[$a_{\rm nd} > 10$\,nm, e.g. see Fig. \ref{fig_spect_nanod2},][]{1999ApJ...521L.133G,2002A&A...384..568V,2004ApJ...614L.129H,2009ApJ...693..610G}, it does not yet appear necessary to enter into the complexities of the thermal processing of temperature-fluctuating, sub-10\,nm nano-particles. Thus, the assumption of thermal equilibrium nano-diamond temperatures close to hot stars appears to be good for now.

\begin{table*}
\caption{
Nano-diamond, a-C, and a-C:H temperatures at 10\,{\tiny AU} as a function of particle radius, $a_{\rm d}$:   
for HR\,4049 the dust temperatures are at 50\,{\tiny AU}$^\star$.}
\centering                          
\begin{tabular}{l c c c c c c}        
\hline\hline \\[-0.25 cm]                 
object & \multicolumn{6}{c}{$T_{\rm d}$\,[ K ]}   \\    
$a_{\rm d}$ [ nm ]  & 0.5 & 1 & 3 & 10 & 30 & 100  \\    
\hline \\[-0.25 cm]                        
 & \multicolumn{6}{l}{nano-diamonds}   \\[0.1cm]  

HR\,4049$^\star$   &  476-687  &   549--767  &   714--1094  &   973--\underline{1656}  &  \underline{1769}--{\bf 2200}  &   \underline{1393}--\underline{1917}   \\ 
    
Elias\,1      &   441--593  &   505--656  &   649--912  &  878--\underline{1430}  &  \underline{1517}--{\bf 2060} & 1189--\underline{1624}  \\     
HD\,97048 &  532--890  &   626--1013  &  844--\underline{1523}  & \underline{1240}--{\bf 2052}  &  {\bf 2152}--({\bf \underline{2395}})  & 1089--({\bf \underline{2346}}) \\[0.1cm]  
 & \multicolumn{6}{l}{amorphous carbon: a-C [ $E_{\rm g} = 0.1$\,eV ] } \\[0.1cm]  

HR\,4049$^\star$   &  1351  &   969  &   728  &   681  &  687  &   648   \\  
 
Elias\,1      &   1114  &   808  &   579  &  535  &  538  &  494  \\     
HD\,97048 &  1273  &   884  &  634  &  584  & 584  &  524  \\[0.1cm]  
 & \multicolumn{6}{l}{hydrogenated amorphous carbon: a-C:H [ $ E_{\rm g} = 2.67$\,eV ] }  \\[0.1cm]  

HR\,4049$^\star$   &  731  &   853  &   950  &   995  &  1089  &   1150   \\      
    
Elias\,1      &   640  &   731  &   803  &  837  &  901  &  899  \\     
HD\,97048 &  735  &   858  &  958  &  1009  & 1088  &  1035  \\[0.1cm]  
\hline    
\end{tabular}
\tablefoot{ 
\underline{Underlined} temperatures , in {\bf bold face} and [{\bf \underline{underlined bold face}}] indicate where the nano-diamonds are likely to be \underline{dehydrogenated}, {\bf surface-reconstructed to sp$^2$} or ({\bf \underline{destroyed by sublimation}}), respectively, under the local conditions in the given object.}
\label{tab_Tnd}
\end{table*}

Before proceeding further it is worth underlining the significant differences between the circumstellar environments of the Herbig Ae/Be stars Elias\,1 and HD\,97048 and the evolved binary system HR\,4049. In particular, the much larger stellar radius of the primary in the latter case (see Table \ref{tab_stars}) results in a local radiation field that is orders of magnitude more intense. The model results in Figs. \ref{fig_ndT_profiles} and \ref{fig_aCH_profiles} and Table \ref{tab_Tnd} indicate higher dust temperatures and would therefore predict displaced and/or more extended dust emission in the case of HR\,4049. Note that in this case the model results given in Table \ref{tab_Tnd} are for dust five times more distant than for the Herbig Ae/Be systems, that is at 50\,{\tiny AU} as opposed to 10\,{\tiny AU}. It is worth pointing out that for HR\,4049 we find large ($a_{\rm nd} = 100$\,nm), a-C (amorphous carbon) grain temperatures of $\simeq 900-1100$\,K at 15\,{\tiny AU}, which are in good agreement with the results of \cite{Acke:2013cn} as displayed in their Fig. 13. This result would then appear to be consistent with the large stellar radius ($R_\star = 31\pm9\,R_\odot$) and therefore with the derived distance to this stellar system \citep[$640\pm190$\,pc,][]{Acke:2013cn}.   

Interestingly, and perhaps paradoxically, the most resistant nano-diamonds, that is the coolest, in the inner regions are always the smallest nano-diamonds (e.g. $a_{\rm nd} \lesssim 10$\,nm, see the left hand panels in Fig. \ref{fig_ndT_profiles}). However, remember that these same nano-diamonds will have minimum temperatures below their equilibrium temperatures in the intervals between the (E)UV photon absorption events that stochastically-heat them to many hundreds  of degrees Kelvin. Thus, while they are subject to the same sorts of temperatures as the larger nano-diamonds, this only occurs over very short timescales, and they are otherwise cooler than their calculated equilibrium temperatures. The reason that the smaller nano-diamonds show lower thermal equilibrium temperatures is because they are more emissive  in the $1-30\,\mu$m region, where their thermal emission peaks. This is elucidated in the right hand panels in Fig. \ref{fig_ndT_profiles}, which indicate the relative nano-diamond emission per unit grain volume at a distance of  10\,{\tiny AU} from the star or 50\,{\tiny AU} from the star as in the case of HR\,4049.\footnote{The emission per unit grain volume is here estimated as $B_\lambda(T_{\rm nd,10AU}) / a_{\rm nd}^3$, multiplied by the absorption efficiency, $\langle Q_{\rm abs} \rangle$, averaged over the wavelength range $0.1-1\,\mu$m at the peak of the stellar emission, the latter in order to take into account the size-dependent stellar photon absorption efficiencies.} This shows that smaller nano-diamonds are more emissive, per unit volume, and therefore significantly cooler, than larger nano-diamonds:  stochastically-varying temperatures do not alter this behaviour.

The hottest nano-diamonds in this equilibrium temperature modelling are those with $a = 30$\,nm in all sources. The reason that the $a = 100$\,nm nano-diamonds are not hotter is not due to their emissivity, which is the lowest of all the sizes considered, but rather due to the fact that they absorb less energy from the stellar radiation field because they exhibit lower values of $Q_{\rm abs}$ at wavelengths close to the peak of the stellar emission $\lambda \sim 0.3\,\mu$m \citep[see Table \ref{tab_stars} and Fig. 5 in ][]{2020_Jones_nd_ns_and_ks}. 
For reference the typical temperature for the onset of nano-diamond dehydrogenation is $T_{\rm -H} \sim 1200-1500$\,K \citep{FT9938903635} and that for surface (and bulk) diamond reconstruction to sp$^2$ aromatic structures, that is graphitisation, is $T_{\rm recon} \sim 2000-2100$\,K \citep{Howes_1962,Fedoseev_etal_1986}. These critical temperatures are indicated in the left hand panels of Fig. \ref{fig_ndT_profiles} and reflected in the highlighted nano-diamond temperatures in Table~\ref{tab_Tnd}. 

Once the nano-diamonds are no longer so, that is they have been graphitised, at temperatures above the graphitisation limit ($T_{\rm nd} > 2000$\,K) they will undergo sublimation at rates that will depend upon the particle size. We can estimate their lifetimes in these excited regions by considering them as graphitic.  
The sublimation rate of graphite is of the order of $1-10$\,mg/h at $\simeq$\,2500\,K \citep{Darken_Gurry_1953} and was  measured to be $5.82$\,mg/h in the laboratory at 2350\,K \citep{Tsai_etal_2005}, which is equivalent to $\tau_{\rm d} = 0.77$\,g\,h$^{-1}$\,m$^{-2}$. Translating this into appropriate astronomically-useful numbers we find that this is equivalent to a nano-diamond sublimation rate of 
\begin{equation}
\tau_{\rm sub} = \tau_{\rm d} \ \frac{4 \pi a_{\rm nd}^2 }{m_{\rm C}} \ \approx 4 \times 10^9 \, \left( \frac{ a_{\rm nd} }{ [{\rm 1\,nm}] } \right)^2 \ {\rm atoms \ yr}^{-1} 
\end{equation}
at sustained temperatures of 2350\,K. For the radii that we consider here, that is 0.5, 1, 3, 10, 30, and 100\,nm, with $N_{\rm C} \sim 80$, 600, $2 \times 10^4$, $7 \times 10^5$, $2 \times 10^7$, and $7 \times 10^8$, respectively, this implies very short lifetimes against sublimation, of the order of  only seconds for the smallest nano-diamonds and up to ten minutes for the largest, at $T_{\rm nd} = 2350$\,K. This critical sublimation temperature is shown in Fig. \ref{fig_ndT_profiles} and these sublimation rates only apply to nano-diamonds that remain close to the star ($< 10-50${\tiny AU}) and that are maintained at temperatures $\geqslant 2350$\,K, that is for the largest nano-diamonds closest to the hottest stars (see Table~\ref{tab_Tnd}).\footnote{N.B., For smaller nano-diamonds the sublimation rate should be integrated over the stochastic temperature fluctuations about the thermal equilibrium temperature and therefore the lifetime derived here may be an over- or an under-estimate but is, nevertheless, probably a good order of magnitude estimate.}

The thermal processing of nano-diamonds in intense radiation field environments is therefore likely to proceed something along the lines of the following sequence:  
\vspace*{0.2cm} \\
\hspace*{1.1cm}  {\tiny dehydrogenation \hspace*{0.4cm} graphitisation \hspace*{0.6cm} sublimation} \\
\hspace*{0.5cm} H-nd \hspace*{0.5cm} $\rightarrow$ \hspace*{0.5cm}  -nd \hspace*{0.5cm}  $\rightarrow$ \hspace*{0.5cm}  a-C \hspace*{0.5cm}  $\rightarrow$ \hspace*{0.5cm}  a-C$_{\rm (g)}^\uparrow$ \\ 
\hspace*{1.3cm}  {\tiny 1200-1500\,K \hspace*{0.60cm} 2000-2100\,K \hspace*{0.75cm} $\gtrsim 2350$\,K } \\ 
\vspace*{-0.1cm} \\
where H-nd and -nd indicate hydrogenated and dehydrogenated nano-diamond, respectively, a-C indicates graphitised nano-diamond and a-C$_{\rm (g)}^\uparrow$ sublimating, graphitised, nano-diamond. The final sublimation step proceeds on timescales of only seconds to minutes at temperatures in excess of $\approx 2500$\,K and in the inner regions of proto-planetary discs graphitised, nano-diamond destruction would seem to be unavoidable and essentially instantaneous. Paradoxically, if nano-diamond emission is observed as close as $10-50$\,{\tiny AU} to a central star then the nano-diamonds located there must, seemingly, have a fleeting existence in the intense radiation field. 

\begin{figure*}
\centering
\begin{center} $
\begin{array}{cc}
   \includegraphics[width=9.0cm]{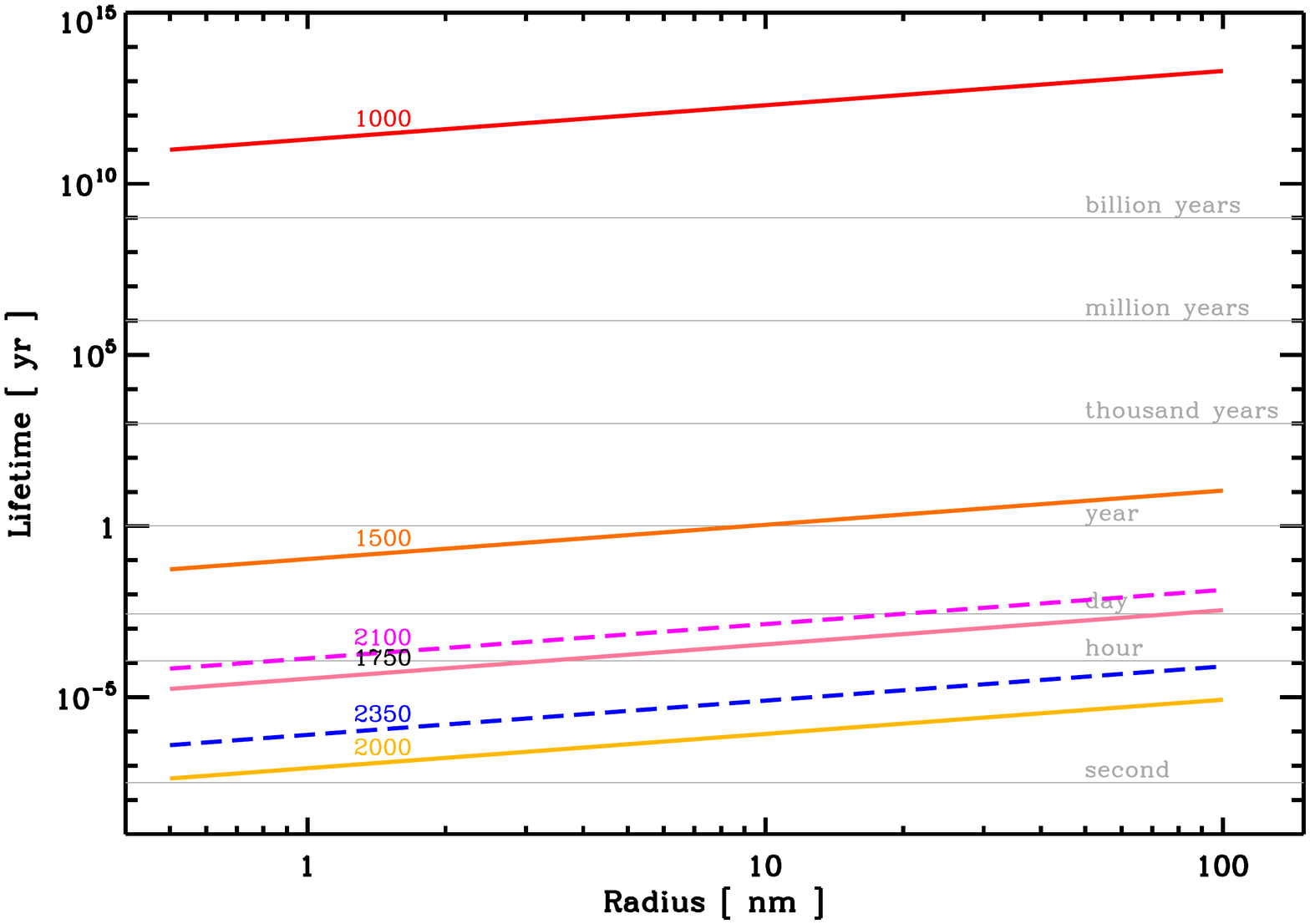} 
   \includegraphics[width=9.0cm]{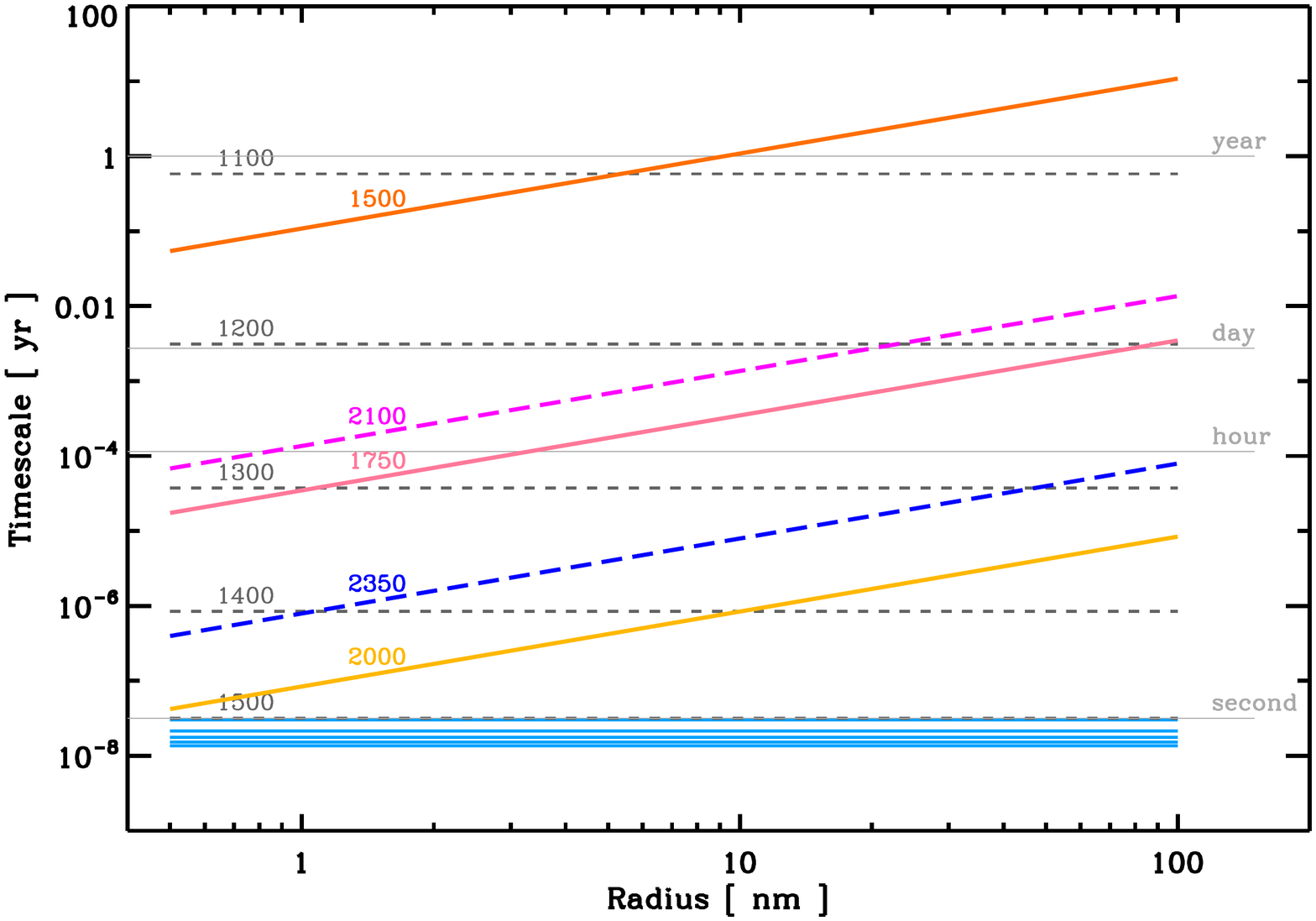}
\end{array} $
\end{center}
\caption{Nano-diamond lifetimes (left panel) as a function of radius and temperature: 1000, 1500, 1750, 2000, 2100, and 2350\,K, red, orange, pink, yellow, violet, and blue, respectively. The solid lines correspond to sublimating nano-diamond and the dashed lines to sublimating nano-diamond transformed into graphite. The right panel shows a zoomed view and also indicates the dehydrogenation timescales (grey dashed lines) at nano-diamond temperatures $T_{\rm nd} = 1100$, 1200, 1300, 1400, and 1500\,K. The light blue lines show the re-hydrogenation timescale for a gas density $n_{\rm H} = 10^6$\,cm$^{-3}$ at temperatures $T_{\rm gas} = 200$, 400, 600, 800, and 1000\,K (upper to lower).}
\label{fig_ndT_lifetimes}
\centering
\end{figure*}

As evaporation does not start at the boiling point, so sublimation has no strict onset temperature as implied above. Hence, we need to consider nano-diamond sublimation in more detail and, in particular, to determine the size- and temperature-dependent rates. To this end we adopt and adapt the Arrhenius equation methodology of \cite{Kobayashi:2009kw} who give the mass loss rate due to sublimation from a carbon grain at a temperature $T_{\rm nd}$ as:
\begin{equation}
-\frac{dm}{dt} = \pi \, a_{\rm nd}^2 \ P_v(T) \ \left( \frac{m_{\rm C}}{2 \, \pi \, k_{\rm B} \, T_{\rm nd}} \right)^{0.5}
\end{equation}
where $m_{\rm C}$ is the mass of a carbon atom sublimating from the grain surface, $k_{\rm B}$ is the Boltzmann constant and the saturated vapour pressure term, $P_v(T)$, is given by 
\begin{equation}
P_v(T) = P_0(T) \ {\rm exp}\left( \frac{m_{\rm C} \, H_{\rm sub}}{k_{\rm B} \, T_{\rm nd}} \right)
\end{equation}
with $P_0(T) =  4.31 \times 10^{16}$\,dyne\,cm$^{-2}$ and the enthalpy of sublimation $H_{\rm sub} = 7.27 \times 10^{11}$\,erg\,g$^{-1}$ taken to be that of graphite \citep{Kobayashi:2009kw}, which we adopt for $T_{\rm nd} > 2050$\,K assuming that the nano-diamonds have been completely graphitised.\footnote{In this case we adjust the graphitised grain radius to conserve the particle mass because of the lower density of graphite (2.24\,g\,cm$^{-3}$) compared to diamond (3.52\,g\,cm$^{-3}$).} For $T_{\rm nd} \leqslant 2050$\,K the nano-diamond bulk remains as such and in this case the removal of a carbon atom is equivalent to breaking four half bonds per carbon atom.\footnote{Each carbon atom is tetrahedrally-coordinated with and shares sp${^3}$ covalent bonds, each of bond energy 3.65\,eV, with four neighbouring carbon atoms. The breaking of four half bonds is is equivalent to losing de-hydrogenated edge-type surface carbons, that is atoms that have lost two hydrogens.} This requires $2 \times 3.65$\,eV $= 7.3$\,eV per atom and equates to an enthalpy of sublimation for nano-diamonds of $7.3 \times 1.602 \times 10^{-12} \times {\rm N_{Avo}} / 12 = 5.87 \times 10^{11}$\,erg\,g$^{-1}$, where N$_{\rm Avo}$ is Avogadro's number. Thus, the enthalpies of sublimation for graphite and diamond estimated here are not too dissimilar but, given that they enter into an exponential term the small difference becomes important. In the left hand panel of Fig.~\ref{fig_ndT_lifetimes} we plot the nano-diamond lifetimes as a function of radius for a some typical temperatures, that is  1000, 1500, 1750, 2000, 2100, and 2350\,K.
\footnote{N.B. In Fig.~\ref{fig_ndT_lifetimes} temperatures lower than 1000\,K are not considered because they result in lifetimes and timescales that are too long to be of any significance in the considered regions.} Clearly lifetimes increase with size (mass) because larger grains contain more atoms and take longer to sublimate. Interestingly this plot also shows that the transformation of diamond to graphite affords the grains some protection due to the higher carbon-carbon bond energies in graphite. This is indicated by the more than two orders of magnitude increase in the grain lifetimes between 2000\,K (solid yellow line: nano-diamond) and 2100\,K (dashed violet line: graphite). This is also apparent from the generally lower temperatures of a-C grains with respect to nano-diamonds, as can be seen by comparing their respective temperature profiles in Figs.~\ref{fig_ndT_profiles} and  \ref{fig_aCH_profiles} and their temperatures at 10 or 50\,{\tiny AU} as given in Table~\ref{tab_Tnd}.

The nano-diamond lifetimes determined here (seconds to minutes) are in broad agreement with the sublimation times at the critical sublimation limit ($2350$\,K) given above and based on graphite sublimation studies \citep{Darken_Gurry_1953,Tsai_etal_2005}. However, as we shall see, any uncertainties in the adopted parameters are not significant because the derived sublimation destruction timescales are short compared to any relevant dynamical timescales.

\section{Nano-diamond survival near hot stars}
\label{sect_recon}

It is perhaps obvious that the degree of surface (de)hydrogenation and bulk/surface reconstruction to sp$^2$ (graphitisation) will critically affect the optical properties of nano-diamonds and hence their temperatures in a given environment. Fully hydrogenated nano-diamonds, with or without a disordered bulk, exhibit more IR absorption modes and will be cooler because of increased emission, with respect to dehydrogenated nano-diamonds with pristine  bulk properties. Fully-dehydrogenated nano-diamonds will clearly show no CH band emission and likely show temperatures closer to that of the largest nano-diamonds (see Fig. \ref{fig_ndT_profiles}), consequently they are less emissive and naturally hotter. Partially-dehydrogenated nano-diamonds will exhibit intermediate spectral properties and  temperatures. 

From the nano-diamond temperature profiles in Fig. \ref{fig_ndT_profiles}, and the synthesis presented in Table \ref{tab_Tnd}, it can clearly be seen that,  close to hot stars ($R \simeq 10$ or 50\,{\tiny AU}),  some sizes of nano-diamonds would apparently exceed the temperature stability range for fully hydrogenated particles. This implies that if they are present in such regions they would, seemingly, not exhibit any observable $3-4\,\mu$m CH$_n$ modes but could still contribute to the continuum emission at these short wavelengths. In general, and close to the stars ($\lesssim 10-40$\,{\tiny AU} and out to 150\,{\tiny AU} in the case of HR\,4049) in these proto-planetary discs the larger nano-diamonds ($a \gtrsim 10$\,nm) exceed the dehydrogenation temperature limit ($T_{\rm -H} \gtrsim 1200-1500$\,K) and in a few cases (for distances $< 5-20$\,{\tiny AU}, out to 70\,{\tiny AU} for HR\,4049) grains with radii $a \gtrsim 10$\,nm also exceed the temperature limit for reconstruction to sp$^2$, that is graphitisation ($T_{\rm recon} \gtrsim 2000-2100$\,K). Further, those with temperatures in excess of $\sim$\,2500\,K will, as shown above, sublime on timescales of the order of only seconds to minutes. Note that the source HR\,4049 represents the most extreme of the three cases for the survivability of nano-diamonds, it also the object that appears to lack an x-ray source and that exhibits a $3-4\,\mu$m spectrum lacking sub-structure in the 3.43 and $3.53\,\mu$m bands. 

These results are somewhat dependent upon the state of the nano-diamonds for the six sets of nano-diamond optical properties considered here (i.e. for un-irradiated or irradiated grains $f_{\rm H} = 0, 0.25$ and 1). However, the derived nano-diamond temperatures are more sensitive to the degree of surface hydrogenation than to the material irradiation state because the strongest emission, and therefore the dominant cooling, is principally through CH$_n$ bands in the $3-4\,\mu$m region (see the right hand panels in Fig. \ref{fig_ndT_profiles}). We conclude that nano-diamond temperatures, at any given distance from the star, increase markedly with decreasing surface hydrogenation. Further, their temperatures are slightly lower when the particle bulk material is in an irradiated state because of the addition of modes in the $6-30\,\mu$m region that aid cooling, this result is likely to hold true for any bulk disorder or heteroatoms that induce long wavelength modes. 

The nano-diamond temperature calculations presented here are obviously based upon a simplified and idealised picture of the nano-diamond properties in that they evidently do not self-consistently take into account the evolution of the material structure and its optical properties. However, the addition of olefinic and aromatic C$-$C bands into the scheme of things (in the $\lambda = 6.1-6.7\,\mu$m region)  will not change the results significantly because their modes are, in general, about an order of magnitude weaker than the predominant CH$_n$ modes (be they of sp$^2$ or sp$^3$ origin) and have integrated cross-sections similar to those of aliphatic C$-$C bands \citep[see Table 2,][]{2020_Jones_nd_ns_and_ks}.

Nevertheless, a composition/size selection effect could operate because larger nano-diamonds are eroded down to smaller sizes, through sublimation/vaporisation, and may preserve some vestige of surface hydrogenation and/or, depending on the local physical conditions (gas density and temperature), they may be re-hydrogenated, more emissive and therefore cooler. Thus, if such an effect operates it would favour the survival of the smallest, surface-hydrogenated nano-diamonds. For instance the temperatures of dehydrogenated/reconstructed nano-diamonds at a given distance from the star will likely be similar to or greater than those for the 30\,nm radius particles because surface reconstruction to sp$^2$ carbon will preserve strong absorption in the $0.2-2\,\mu$m wavelength region where the peak of the stellar emission occurs. The grains would therefore not be saved by a reduced absorption effect at these wavelengths, which is the explanation for the lower temperatures of the 100\,nm radius (de)hydrogenated nano-diamonds compared to 30\,nm nano-diamonds. Further, nano-diamonds with radii $\gg 30$\,nm cannot be saved either because as they undergo dehydrogenation/reconstruction their absorption will increase in the $0.2-2\,\mu$m wavelength region causing them to heat even further. Nano-diamonds could therefore become trapped within destructive feedback loops, whereby dehydrogenation leads to increased temperatures leading to increased erosion rates due to sublimation/evaporation at temperatures $\geqslant 2500$\,K (see Fig.~\ref{fig_ndT_lifetimes}). 

For comparison with the nano-diamond results we also show,  in Fig.~\ref{fig_aCH_profiles}, the temperature profiles of same-size a-C and a-C:H  particles, with $E_{\rm g} = 0.1$ and 2.67\,eV, respectively, for the same three objects. What we see in this figure and in Table \ref{tab_Tnd} is that the a-C(:H) grains, whether fully hydrogenated (a-C:H) or hydrogen-poor (a-C), are generally cooler than the equivalent size nano-diamonds.   In $10-30$\,{\tiny AU} regions of the two Herbig Ae/Be stars the a-C(:H) grain temperatures range $\sim 300-1100$\,K versus $\sim 300-2400$\,K for nano-diamonds of the same sizes ($a = 0.5-100\,\mu$m). This is primarily due to the enhanced hydrogen atom fraction and therefore stronger emission in the CH$_n$ bands in the case of a-C:H and in a-C grains is due to stronger continuum emission, both effects enhanced with respect to nano-diamond optical properties. While this may perhaps indicate the possible survival of a-C(:H) grains close to hot stars it should be pointed out that a-C(:H) particles are more fragile than nano-diamonds and will succumb to photo-fragmentation(-dissociation) processes \citep[e.g.][]{2013A&A...558A..62J,2014A&A...569A.119A,2015A&A...581A..92J,2015A&A...584A.123A,2019A&A...569A.100B,2020A&A...639A.144S,2021A&A...649A.148S}. In the case of HR\,4049 the a-C(:H) grain temperatures are obviously significantly higher in these inner regions ($10-30$\,{\tiny AU}) but, as pointed out above, these large amorphous carbon grain temperatures are consistent with the results of \cite{Acke:2013cn}.

With the above rather short timescales in mind we now need to delve into some of the other characteristic timescales in order to see whether they may be mitigated by other processes. For instance, dehydrogenation can be counteracted or even nullified by incident H atom collision, sticking, and surface re-hydrogenation. Also, it is possible that radiation pressure can move nano-diamond out of harm's way. We now consider these effects. 

The right hand panel of Fig.~\ref{fig_ndT_lifetimes} again plots the nano-diamond lifetimes, for temperatures 1500, 1750, 2000, 2100, and 2350\,K, and compares these with the dehydrogenation timescales (grey dashed lines) at nano-diamond temperatures $T_{\rm nd} = 1100$, 1200, 1300, 1400, and 1500\,K. Lower temperatures are not considered because the timescales are too long to be of significance and densities $> 10^6$\,cm$^{-3}$ lead to sub-second rehydrogenation timescales.  In the absence of experimental data for hydrogenated nano-diamonds, the H loss timescales, $\tau_{\rm -H}$, were estimated using an Arrhenius Equation approach, that is,
\begin{equation}
\tau_{\rm -H} = \frac{ 4 \pi \, a_{\rm nd}^2 \, m_{\rm H} / \sigma_{\rm H} }{ 4 \pi \, a_{\rm nd}^2 \ P_{\rm v,H}(T_{\rm nd}) \ \surd [ m_{\rm H} / ( 2 \pi \, k_{\rm B} \, T_{\rm nd} ] ) },  
\end{equation}
where the numerator is the mass of H atoms per nano-diamond surface and the denominator is the H atom mass loss rate from the surface, where 
\begin{equation}
P_{\rm v,H}(T_{\rm nd}) = A_{\rm H} \, {\rm exp} \{ E_{\rm act}({\rm H}) \, / \, ( k_{\rm B} \, T_{\rm nd} ) \}
\end{equation}
is the equivalent of a saturated vapour pressure for atomic hydrogen evaporating from a nano-diamond surface. The area per surface CH bond, $\sigma_{\rm H}$, is $\simeq 0.05$\,nm$^2$ \cite[an average of the values for CH and CH$_2$ derived in][]{2020_Jones_nd_CHn_ratios}. The activation energy for surface H loss $E_{\rm act}({\rm H})$ is set equal to 5\,eV, that is somewhat larger that the typical aliphatic CH bond energies of $4.3-4.8$\,eV to allow an excess of energy for H atom removal from the surface following CH bond breaking. The assumed pre-exponential factor, $A_{\rm H} = 4.23 \times 10^{17}$\,dyne\,cm$^{-2}$, sets dehydrogenation lifetimes of 1 day at 1200\,K and 1\,s at 1500\,K, which seem reasonable given that the typical dehydrogenated temperature range appears to be $1200-1500$\,K. These timescales are clearly best-guess estimates in the absence of appropriate data.

The counteracting nano-diamond re-hydrogenation or H atom attachment rate can then be estimated via 
\begin{equation}
\tau_{\rm H} = \Bigg\{  n_{\rm H} \ \pi \, a_{\rm nd}^2 \, \left( \frac{ 2 \, k_{\rm B} \, T_{\rm gas} }{m_{\rm H} } \right)^{\frac{1}{2}} \ s_{\rm H}  \Bigg\}^{-1}
\end{equation}
where $n_{\rm H}$ and $T_{\rm gas}$ are the gas density and temperature, respectively, and $s_{\rm H}$ is the collision, sticking, and re-hydrogenation efficiency. As a conservative value we here take $s_{\rm H}= 0.01$, that is 1\% of colliding H atoms lead to surface re-hydrogenation, and implicitly assume that some fraction of the hydrogen in the gas is in atomic form. The light blue lines in the right panel of Fig.~\ref{fig_ndT_lifetimes} show the re-hydrogenation timescales for a gas  temperatures $T_{\rm gas} = 200$, 400, 600, 800, and 1000\,K at a gas density $n_{\rm H} = 10^6$\,cm$^{-3}$. This figure shows that, for gas densities  $> 10^6$\,cm$^{-3}$ and $T_{\rm gas} > 200$\,K, the nano-diamond re-hydrogenation timescales are significantly shorter than that for H loss by sublimation from the surface, a result that is independent of particle size. Thus, with our estimated parameters and reasonable assumptions about the local physical conditions, it appears that nano-diamond surface re-hydrogenation can counteract H atom loss through sublimation. Nevertheless, it is clear that even though the timescales for nano-diamond sublimation are significantly longer than those for re-hydrogenation the sublimation timescales are too short to ensure long term nano-diamond survival close to hot stars.   

\begin{figure}
\centering
   \includegraphics[width=9.5cm]{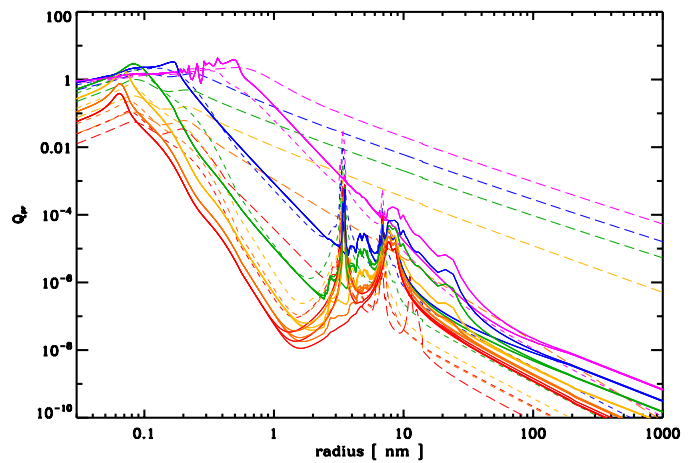}
      \caption{Nano-diamond and a-C(:H) material radiation pressure efficiency factors $Q_{\rm pr}$ for nano-diamonds (solid), aliphatic-rich a-C:H grains with $E_{\rm g} = 2.67$\,eV (thin short-dashed lines) and aromatic-rich a-C grains with $E_{\rm g} = 0.1$\,eV (thin short-dashed lines). The line colour coding is the same as in Fig. \ref{fig_spect_nanod2}.}
      \label{fig_nanod_Qs}
\end{figure}

\begin{figure}
\centering
   \includegraphics[width=9.5cm]{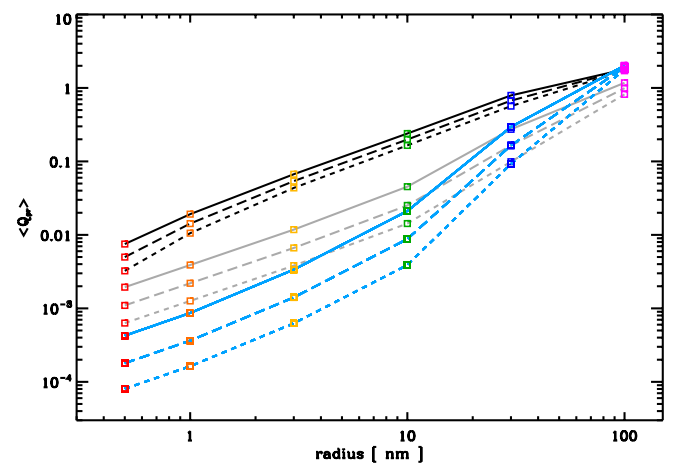}
      \caption{Nano-diamond and a-C(:H) material Planck-averaged radiation pressure efficiency factors $\langle Q_{\rm pr}\rangle$ for nano-diamonds (cobalt), aliphatic-rich a-C:H grains with $E_{\rm g} = 2.67$\,eV (grey) and aromatic-rich a-C grains with $E_{\rm g} = 0.1$\,eV (black) 10\,{\tiny AU} from HR 4049 (short-dashed), Elias\,1 (long-dashed), and HD\,97048 (solid). The line colour coding is the same as in Fig. \ref{fig_spect_nanod2}.}
      \label{fig_nanod_QPave}
\end{figure}

\begin{figure}
\centering
   \includegraphics[width=9.5cm]{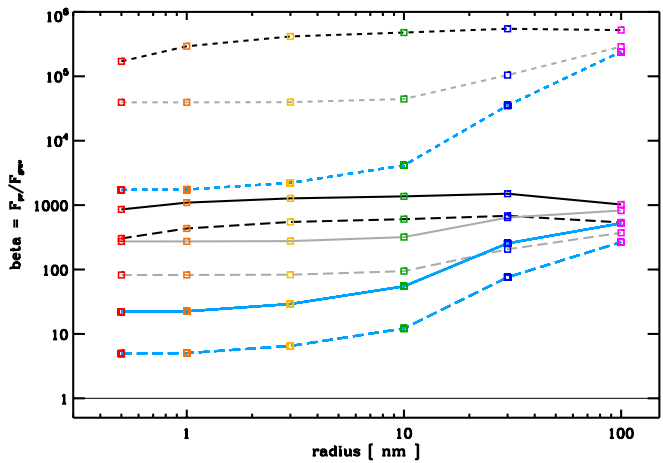}
      \caption{Nano-diamond and a-C(:H) material beta factors, $\beta = F_{\rm pr}/F_{\rm grav}$, the ratio of the radiation pressure and gravitational forces for nano-diamonds (cobalt), aliphatic-rich a-C:H grains with $E_{\rm g} = 2.67$\,eV (grey) and aromatic-rich a-C grains with $E_{\rm g} = 0.1$\,eV (black) 10\,{\tiny AU} from HR 4049 (short-dashed), Elias\,1 (long-dashed), and HD\,97048 (solid). The data point colour coding is for the same sizes as in Fig. \ref{fig_spect_nanod2}.}
      \label{fig_nanod_betas}
\end{figure}

We now turn our attention to the possible nano-diamond saving graces of radiation pressure and outward drift. To this end we have calculated the radiation pressure efficiency factors $Q_{\rm pr}$ for nano-diamonds and a-C(:H) materials (aliphatic-rich a-C:H grains with $E_{\rm g} = 2.67$\,eV and aromatic-rich a-C grains with $E_{\rm g} = 0.1$\,eV), which are plotted in Fig.~\ref{fig_nanod_Qs}. From these we calculate their Planck-averaged radiation pressure efficiency factors $\langle Q_{\rm pr}\rangle$ 10\,{\tiny AU} from the stars HR 4049, Elias\,1, and HD\,97048 (see Fig.~\ref{fig_nanod_QPave}), analogous to the calculation of $\langle Q_{\rm abs}\rangle$ (see Section~\ref{sect_Ts}) but with $Q_{\rm abs}$ replaced by $Q_{\rm pr}$. The radiation pressure force, $F_{\rm pr}$, and gravitational force, $F_{\rm grav}$, acting on nano-diamond close to stars are, respectively, 
\begin{equation}
F_{\rm pr} = \pi \ \frac{R_\star^2}{d^2} \ \frac{1}{c} \ \pi \, a_{\rm nd}^2 \, \langle Q_{\rm pr} \rangle \ 4 \, \sigma \, T_{\rm eff}^4 
\end{equation}
and
\begin{equation}
F_{\rm grav} = \frac{ {\rm G} \, M_\star }{ d^2 } \ \frac{ 4 \, \pi \, a_{\rm nd}^3 }{ 3 } \ \rho_{\rm nd} . 
\end{equation}
where $R_\star$ and $M_\star$ are the stellar radius and mass, $d$ the distance from the star, G the gravitational constant,  $\sigma$ the Stephan-Boltzmann constant and $\rho_{\rm nd}$ the nano-diamond specific mass density (3.52\,g\,cm$^{-3}$). The factor, $\beta$, is the ratio of these two forces, that is 
\begin{equation} 
\beta = \frac{ F_{\rm pr} }{ F_{\rm grav} } = \left( \frac{3 \, \pi \, \sigma}{c \, {\rm G}} \right) \ \left( \frac{R_\star^2 \, T_{\rm eff}^4}{M_\star} \right) \ \left( \frac{\langle Q_{\rm pr} \rangle}{a_{\rm nd} \, \rho_{\rm nd}} \right) ,  
\label{eq_beta}
\end{equation}
for grains with $\beta < 1$ gravity dominates, in rare cases $\beta = 1$ and the two forces are in equilibrium, otherwise $\beta > 1$ and radiation pressure wins out. We note that the ratio parameter $\beta$ is dimensionless and independent of distance. The nano-diamond and a-C(:H) material $\beta$ values in the vicinity of HR 4049, Elias\,1, and HD\,97048 are plotted in Fig.~\ref{fig_nanod_betas} and show that, for all the materials considered here, $\beta > 1$ and up to three orders of magnitude greater in the case of the absorbing, aromatic-rich a-C particles. Note that in HR\,4049 the larger $\beta$ values for the a-C grains ($\lesssim 10^6$) are driven by the stellar radius, all other terms being not too dissimilar to those for Elias\,1 and HD\,97048 (e.g. see Eq. \ref{eq_beta}). Note that the principal size-for-size difference between a-C(:H) dust \cite[$1.3-1.6$\,g\,cm$^{-3}$,][]{2012A&A...542A..98J} and nano-diamonds is driven by the material specific density differences and, hence, the effect of gravity is more marked on the denser nano-diamonds. Thus, in all cases, nano-diamonds and carbonaceous grains should be expelled from the inner disc regions of the three considered stars.

\begin{figure}
\centering
   \includegraphics[width=9.5cm]{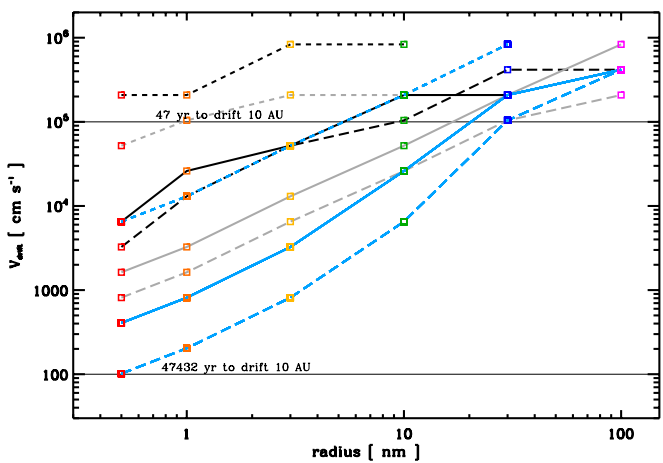}
      \caption{The nano-diamond and a-C(:H) material drift velocities (for $n_{\rm H} = 10^8$\,cm$^{-3}$ and $T_{\rm gas} = 500$\,K) for nano-diamonds (cobalt), aliphatic-rich a-C:H grains with $E_{\rm g} = 2.67$\,eV (grey) and aromatic-rich a-C grains with $E_{\rm g} = 0.1$\,eV (black) 10\,{\tiny AU} from HR 4049 (short-dashed), Elias\,1 (long-dashed), and HD\,97048 (solid). The data point colour coding is for the same sizes as in Fig. \ref{fig_spect_nanod2}.}
      \label{fig_nanod_Vdrift}
\end{figure}

However, the $\beta$ factor takes no account of the dynamical aspects because the outward radiation pressure on the dust is opposed by collisional drag with the gas. Following \cite{1987ApJ...318..674M} and \cite{1996ApJ...469..740J} the drag force acting on a dust particle, $F_{\rm drag}$, is given by 
\begin{equation}
F_{\rm drag} = \pi \, a_{\rm nd}^2 \ V_{\rm drift} \ 1.4 \, m_{\rm H} \, n_{\rm H} \ \left( \ V_{\rm drift}^2 + \frac{ 128\, k_{\rm B} T_{\rm gas} }{ 9 \, \pi \, 1.4 \, m_{\rm H} } \ \right)^\frac{1}{2}
\end{equation}
where $V_{\rm drift}$ is the drift velocity when the forces acting on the particle are in equilibrium. The dust drift velocities, in a gas of density $n_{\rm H}$ and temperature $T_{\rm gas}$ at a distance, $d$, from the star, are then obtained by solving the following equation for $V_{\rm drift}$ 
\begin{equation}
F_{\rm pr}(d)  = F_{\rm drag}(V_{\rm drift},d) + F_{\rm grav}(d)
\label{eq_vdrift}
\end{equation}
for given grain radii and material properties. In Fig.~\ref{fig_nanod_Vdrift} we show the nano-diamond and a-C(:H) material drift velocities at 10\,{\tiny AU} calculated by solving Eq~(\ref{eq_vdrift}) for all of our considered materials. This figure also indicates some likely drift timescales and shows that to move the largest and hottest nano-diamonds 10\,{\tiny AU}, that is well out of harm's way of the intense stellar radiation field, requires of the of the order of tens of years for a disc gas density of $10^8$\,cm$^{-3}$, compared to sublimation timescales of the order of seconds to minutes in these same regions. This is also holds for HR\,4049 with its radiation field that is, at a given distance, more than two orders of magnitude larger than is the case for Elias\,1 and HD\,97048. In HR\,4049 at a distance of $\simeq 50$\,{\tiny AU}, where we would perhaps expect the nano-diamonds to survive and emit (see Fig. \ref{fig_ndT_profiles}), distance dilution by a factor of $(10/50)^2 = 0.04$ results in a radiation field at 50\,{\tiny AU} that is only about an order of magnitude larger than for Elias\,1 and HD\,97048 at distances of 10\,{\tiny AU}. Thus, the HR\,4049 dust drift velocities at 50\,{\tiny AU} are about an order of magnitude less than those shown in Fig.~\ref{fig_nanod_Vdrift} and therefore comparable to those for Elias\,1 and HD\,97048 at 10\,{\tiny AU}. This implies that the relevant dynamical timescales are too long to eject and therefore protect nano-diamond from the rapid sublimating conditions within the inner disc regions, that is at distances $\lesssim 10$ or 50\,{\tiny AU}. This leaves us with an obvious paradox in that the inner disc regions have the right conditions to make nano-diamonds shine but are also extremely detrimental to their well-being. 

However, if the nano-diamonds exist further out in the disc, for example at 30\,{\tiny AU}, or 200\,{\tiny AU} in HR\,4049, where the $a = 10-100\,\mu$m nano-diamonds have temperatures of $\sim 500-1000$\,K, $\sim 500-1050$\,K, and $\sim 700-1550$\,K in HR\,4049, Elias\,1 and HD\,97048, respectively, they will be cooler, more stable and survive for considerably longer. We discuss the detailed consequences of this possibility in Section \ref{sect_formation}. 

We note that the dynamical particle drift timescales derived here are probably underestimated because we use the values at  10\,{\tiny AU}  instead of performing a full integration as a function of distance from the stars. Such detailed calculations would seem to be a little premature and we therefore leave them until such time as we can perform them within the context of a 3-D disc model with fully-determined density, temperature, and attenuation dependent radiation field profiles. 

Despite the simplicity of the calculations performed here the order of magnitude estimates that we make leave us with something of a conundrum because they appear to indicate that nano-diamonds cannot survive long in the inner regions where they are emitting brightly and must therefore be transient dust species. It seems that the extreme conditions in the discs where nano-diamonds are seen, that is intense radiation fields, and possibly x-ray flares, are required to form them but that these conditions also sound their death knell. Thus, for nano-diamonds to be observed close to hot stars seems to require a very fine balance between excitation and sublimation, whether or not they are formed in-situ.\footnote{Note that, other than those extracted from primitive meteorites in the solar system, nano-diamonds have never been seen in any other extra-solar environment even though some of them clearly seem to have been associated with distant supernov\ae.} This may perhaps explain why observations of sources exhibiting nano-diamond emission are so rare.

\section{Things of note --- A pre-discussion summary}
\label{sect_notes}

This section enumerates, in a more digestible manner, what seem to be the most notable aspects of this work. 
In the following we consider the fate of nano-diamonds in the regions around hot stars ($T_{\rm eff} \simeq 7,500 - 10,500$\,K) and at distances of the order of $10-50$\,{\tiny AU}, that is environments similar to those in the few objects where they have been observed. Further, we assume optically thin circumstellar regions and only consider nano-diamond equilibrium temperatures, which are more-or-less valid close to hot stars because of the very intense radiation fields.\footnote{This would not be the case in the diffuse ISM where nano-diamonds would be predominantly stochastically-heated.} 

In the following enumerated items we highlight the most interesting and intriguing aspects uncovered in the explorations of the previous sections:

\begin{enumerate} 

\item Close to stars, smaller nano-diamonds ($a \lesssim 30$\,nm) are cooler than their larger cousins. However, larger (nano-)diamonds ($a \sim 100$\,nm), with almost bulk diamond-like properties, will be somewhat cooler than slightly smaller nano-diamonds ($a \sim 30$\,nm) because they absorb less of the stellar UV. 

\item The critical nano-diamond temperatures are:  $\sim 1200-1500$\,K de-hydrogenation, $\sim 2000-2100$\,K graphitisation (sp$^2$ reconstruction), and $\gtrsim 2500$\,K (graphitised) nano-diamond sublimation. 

\item Nano-diamonds with $a \gtrsim 10$\,nm can be dehydrogenated close to stars, those with $a_{\rm nd} > 10$\,nm will be hotter and undergo (surface) reconstruction to sp$^2$ aromatic carbon and those with $a_{\rm nd} \gtrsim 30 $\,nm will be even hotter and sublime. 

\item Dehydrogenation will be counterbalanced by re-hydro\-genation if hydrogen is partially atomic and the gas is dense enough ($n_{\rm H} > 10^6$cm$^{-3}$), this conclusion is essentially independent of the gas temperature. 

\item Only nano-diamonds with $a_{\rm nd} < 10$\,nm can seemingly survive unscathed and those that shine close to hot stars must truly be nano-particles (radii $< 10$\,nm). Nevertheless, it appears that all nano-diamonds will sublime on rather short timescales.  

\item At the same distance from the star a-C(:H) grains, whether of a-C ($E_{\rm g} = 0.1$\,eV) or a-C:H ($E_{\rm g} = 2.67$\,eV), are considerably cooler ($500-1000$\,K)  because they are more emissive than nano-diamonds. 

\item The CH$_n$ band emission of any a-C(:H) grains spatially-associated with nano-diamonds must, mass for mass, come under the observability radar, that is be less abundant than nano-diamonds. However, a-C(:H) materials are considerably more susceptible to photo-fragmentation, photo-processing and photo-dissociation and are themselves unlikely to survive for long in these regions. 

\item The parameter $\beta$, the ratio of the forces of radiation pressure and gravitation acting on dust, is in excess of unity for all of the considered particles. The lowest values of $\beta$ are found for the smallest nano-diamonds ($\beta \simeq 5$ for $a < 10$\,nm). 

\item All particles will experience a net outward force due to radiation pressure. The dispersion in $\beta$ reflects the variation in the Planck-averaged radiation pressure efficiency factors, $\langle Q_{\rm pr} \rangle$. 

\item The derived drift velocities are largest for the largest nano-diamonds, which could remove them from harm's way in tens of years. This same conclusion also holds true for similar size a-C(:H) grains. 

\item However, the largest nano-diamonds likely sublime within only days at temperatures $\simeq 2000$\,K. The derived drift velocities are seemingly too low to protect them from this fate. 

\item Thus, if large nano-diamonds do exist in circumstellar discs they may only do so at large distances from the star, in optically thin regions, or within dense disc regions (where they will be invisible). 

\end{enumerate}

The above model-based scenarios and suppositions need to be tested and verified, or refuted, using a full 3-D disc model including variable optical depths, gas densities, and temperatures. A thorough study of dust evolution, which is coupled to its drift velocity, will obviously require a time-dependent, dynamical model.

\section{Discussion and speculations}
\label{sect_results}

In the highly excited inner regions of proto-planetary discs, that is regions with intense and hard radiation fields, we find that it is the smallest nano-diamonds that will be the most stable against thermal processing leading to the dehydrogenation and re-construction of sp$^3$ surfaces to sp$^2$ carbon. This will occur through the aromatisation of the diamond \{111\}-like and \{100\}-like facets into sp$^2$ de-laminated, aromatic sheets parallel to those surfaces \citep[e.g.][]{1997PhRvB..55.1838Z,Barnard:2005dt}. Such a graphitisation process is observed in the laboratory at temperatures $\gtrsim 2000-2100$\,K \citep{Howes_1962,Fedoseev_etal_1986}. However, it does appear that dehydrogenation could be effectively countered by re-hydrogenation through H atom collision, sticking and reaction on nano-diamond surfaces.  

Of the three sources HR\,4049 appears to be the more extreme. Nevertheless, if we consider its nano-diamond temperatures at 50\,{\tiny AU} rather than 10\,{\tiny AU}, as we will do in the following discussions, we can see that the dust temperatures in the three objects are rather similar (see Table \ref{tab_Tnd}). In this study we have used a blackbody to model the stellar spectral energy distribution of all three stars, which may not reflect the hardness of the radiation field \citep[e.g. see the case of HD\,97048 in][]{2002A&A...384..568V}, we may have underestimated the nano-diamond temperatures, however, this will depend upon the details of the stellar SED and the shape of the nano-diamond optical properties in the EUV.\footnote{Clearly, and well beyond the scope of this article, a much more detailed radiative transfer modelling of these sources is required using a range of dust compositions and a full size distribution in order to fully explore and exploit the potential utility of nano-diamond observations.} In these three objects the nano-diamonds are observed close to the stars where the radiation fields are intense enough that the grains are essentially in thermal equilibrium and the assumption of thermal equilibrium temperatures is therefore justified at this stage.

It is interesting to note that of the three sources considered here HR\,4049, with no known x-ray emission,  shows a $3-4\,\mu$m spectrum that is significantly different from the other two, which exhibit well-defined $3.43$ and $3.53\,\mu$m bands with clear sub-structure \citep[e.g.][]{1999ApJ...521L.133G,2002A&A...384..568V,2004ApJ...614L.129H}. This would appear to indicate that, although extreme environments are indeed required to produce the characteristic (nano)diamond IR spectra, this is not sufficient and x-ray irradiation might also play a role. Thus, there is perhaps a fine balance between their formation, excitation and destruction in extreme sources and that is why they are so rarely observed in astrophysical environments.

The sublimation of the larger nano-diamonds ($a \gtrsim 30$\,nm) close to stars with extremely intense and/or hard radiation fields would therefore provide a natural selection effect for a size-sorting of nano-diamonds, where the coolest and smallest particles preferentially survive.\footnote{The survival of the coolest.} In these highly-excited regions the larger and hotter particles will reconstruct, leading to increased heating but also increased cooling, but not sufficiently so to protect them from  sublimation on timescales of, at most, days. Thus, the evolution of their optical properties through graphitisation leads to the loss of CH modes, that are responsible for a significant degree of the cooling of nano-diamonds, while any increase in the CC modes does not help because they are relatively weak compared to  CH modes. 

In the outer regions of proto-planetary discs, at distances greater than $20-50$\,{\tiny AU} ($150-300$\,{\tiny AU} in HR\,4049), fully-hydrogenated and un-graphitised nano-diamonds of all sizes should be stable against thermal processing. Nevertheless, and given that the presolar nano-diamonds analysed to date all have approximately log-normal size distributions peaking at $\simeq 3$\,nm and extending out to $\sim 10$\,nm \citep{Lewis_etal_1987,1989Natur.339..117L,1996GeCoA..60.4853D}, it seems that nano-diamonds with radii as large as 100\,nm are not the norm. For if they were we would surely see signs of them in the pre-solar grains. A possible explanation for this is that the nano-diamonds are actually formed in the inner regions of proto-planetary discs, by some as yet unspecified process,\footnote{As noted by \cite{1996GeCoA..60.4853D}, the pre-solar nano-diamonds were most probably formed via some vapour phase condensation process.} and that they are there size-sorted as a result of thermal processing, which leads to the observed size distribution biased towards smaller nano-diamonds. However, the flaw in this argument is that the analysed pre-solar nano-diamonds, or at least a large fraction of them, must be of extra-solar origin based upon their anomalous Xe isotopic component (Xe-HL), which is considered characteristic of the nucleosynthetic processes in supernovae \citep{Lewis_etal_1987}. This extra-solar origin is also supported by the fact that they exhibit  $^{15}$N depletions and low C/N ratios that are consistent with carbon-rich stellar environments \citep{1997AIPC..402..567A}. 

There has always been something of a question mark over the origin of the anomalous Xe-HL in pre-solar nano-diamonds. It has been proposed that it arises from implantation \citep[e.g.][]{2000LPI....31.1804V}. However, it is hard to understand how this process could lead to the trapping of Xe atoms in such small particles because incident heavy ions would likely traverse the particle rather that be implanted. If small nano-diamonds are, however, formed by the erosion of much larger particles in intense radiation field environments, it is possible that the heavier Xe atoms could be retained in the grain during downsizing as a result of progressive sublimation. Such an effect would, as required, explain Xe atom trapping in nano-diamonds and also result in a concentration effect that would increase the number of Xe atoms per unit nano-diamond mass. The key flaw in this argument is that at the temperatures required for downsizing by sublimation the nano-diamonds should have been graphitised.

If nano-diamond emission is coming from inner disc regions ($R \sim 10-15$\,{\tiny AU} or $R \sim 50-100$\,{\tiny AU} in the case of HR\,4049) around hot stars, as indeed some fraction of it must, then we seem to have something of a conundrum because the modelling presented here seems to show that they cannot exist there for long and cannot be saved by outward drift due to radiation pressure. This would imply that in the inner disc regions they must be transient species that only shine during their dying throes in extreme environments. If so then these sorts of environment are probably not the principal sites of their formation. However, this conundrum can clearly be lifted if nano-diamond emission also originates from further out in the discs ($R \sim 15-30$\,{\tiny AU}), as indicated by observations \citep[e.g.][]{2004ApJ...614L.129H,2009ApJ...693..610G}. For the nano-diamonds around HR\,4049 this work would predict that their emission must be coming from regions $\sim 100-200$\,{\tiny AU} distant from the primary. In these more distant regions they will be considerably cooler and therefore much longer lived. These possibilities are considered in detail in the following section within the framework of nano-diamond formation and processing.

\section{The formation of nano-diamonds}
\label{sect_formation}

This section briefly summarises some of the observed properties of the nano-diamond emission band and similar sources based on the relevant literature \citep[e.g.][]{2004ApJ...614L.129H,2006A&A...449.1067H,2009ApJ...693..610G,Acke:2013cn,2014ApJ...780...41M,Kokoulina:2021jw}. This is followed with a consideration of the viability of nano-diamond formation within these environments. All of these objects also show the carbonaceous emission bands that are widely observed in the ISM and in PDRs. Given that these bands are actually of mixed aromatic, olefinic, and aliphatic carbonaceous materials the generally-adopted label for them, polycyclic aromatic hydrocarbons (PAHs), is therefore a misnomer. Here they will be referred to as arophatic \citep{2012ApJ...761...35M}, so as to more accurately indicate there origin in a mixed aromatic/olefinic/aliphatic amorphous hydrocarbon material, as per the THEMIS a-C(:H) nano-particles \citep{2012A&A...540A...1J,2012A&A...540A...2J,2012A&A...542A..98J,2013A&A...558A..62J,2017A&A...602A..46J}.

The source characteristics that appear to be required for diamond emission appear to be a hot star ($T_{\rm eff} \sim 7,500-10,500$\,K), a circum(stellar/binary) disc that is optically thick (UV-IR), and a dense disc gas ($n_{\rm H} = 10^8 - 10^{11}$\,cm$^{-3}$) that is most likely carbon rich. Two of the diamond sources are Herbig Ae/Be stars (HD\,97048 and Elias\,1) exhibiting x-ray emission (flares), while HR\,4049, which shows rather featureless diamond bands, is an extreme lambda Boo-type star, with a photosphere depleted in Fe and Si, solar S, C, N and O, and an O-rich outflow. In the latter it is assumed that refractory dust formed within the circumbinary disc resulting in a gas devoid of Fe and Si being accreted onto the primary. Thus, the circumbinary environment of HR\,4049 is dominated by its mass loss history and the dust formed there is freshly-formed stardust. This therefore constitutes proof that nano-diamonds form in-situ in HR\,4049 and indicates that they  are different from the pre-solar nano-diamonds extracted from meteorites.  

It is interesting to note that only $\sim 50$\% of Herbig Ae/Be stars ($T_{\rm eff} \sim 10,000$\,K) show arophatic emission and only $\sim 10$\% show diamond emission. In comparison, diamond emission has not been observed around a T-Tauri star ($T_{\rm eff} < 6,000$\,K) and only $\sim 10$\% of them show arophatic emission bands, likely because the radiation in their circumstellar shells is too weak to excite and/or form arophatic and diamond nano-particles. 

The dust in the diamond sources appears to be principally carbon-rich (exhibiting, variously, diamond, fullerene and arophatic emission bands) and much of the dust likely amorphous carbon. In HR\,4049 the arophatic emission is thought to come more from the bi-polar outflow than from the disc \citep{Acke:2013cn}. The radial sequence of the extent of the dust features with increasing distance from the central star appears to be: IR continuum ($T_{\rm dust} \sim 1000-2000$\,K) $\rightarrow$ (nano-)diamond $\rightarrow$ fullerene / arophatic $\rightarrow$ mm continuum (out to several 100\,{\tiny AU}), with some observed overlap in the emitting regions. This sequence perhaps reflects the effects of decreasing intensity in the dust excitation and hence its evolution with decreasing radiation field. In all cases the diamond emission bands are more extended than the continuum emission but less so than arophatic band emission. Silicate dust emission is not prominent in any of the objects and any silicate/oxide dust component is therefore probably rather well hidden within the denser parts of the discs. 
 
Observations show that the nano-diamond emission originates from the inner disc regions of the Herbig Ae/Be stars, $R \sim 10-30$\,{\tiny AU} with the minimum size of the emitting region  $\sim 10$\,{\tiny AU}, indicating that the nano-diamonds do not exist interior to this limit but do emit in these inner disc regions.\footnote{In Elias~1 the nano-diamond emission column density actually peaks at 30\,{\tiny AU} \citep{2009ApJ...693..610G}.}  Hence, some nano-diamond emission is coming from regions where, perhaps paradoxically, they ought to be rapidly destroyed.  

A viable nano-diamond formation scenario was proposed by \cite{2009ApJ...693..610G}, which posits that diamond nucleates inside bucky onion structures to form nm-sized diamond domains. In this scenario, the nucleated diamond then grows outwards under the effects of (x-ray) irradiation, at optimal dust temperatures of $\sim 600$\,K, with carbon atom sp$^2$ to sp$^3$ conversion progressing outwards until complete transformation of the particle into a $\simeq 100$\,nm nano-diamond has occured \citep{2009ApJ...693..610G}. This mechanism is well supported by the numerous laboratory studies that they cite. As \cite{2009ApJ...693..610G} note the observed anti-correlation between the nano-diamond and arophatic emission bands would appear to support this diamond formation scenario. However, it could be argued that, if diamond forms directly from amorphous hydrocarbonaceous dust, a-C(:H), under the effects of (x-ray) irradiation, then the association between the diamond and arophatic emitting regions ought to be more intimate. A major problem here is that the arophatic band emission comes from nano-particles with far fewer carbon atoms ($\sim 10^3$ atoms for $a \sim 1$\,nm) than the nano-diamonds ($\sim 10^6 - 10^9$ atoms for $a \sim 10 -100$\,nm) and so the arophatic band emitters cannot be the parent particles of the nano-diamonds. We note that in their study of the Herbig Ae/Be star HD\,179218 \cite{Kokoulina:2021jw} find that the inner disc edge ($R \simeq 10$\,{\tiny AU}) is associated with 1700\,K carbonaceous nano-grain emission, which requires replenishment process because of its short survival time there. They estimate that the disk accretion inflow could feed the inner 10\,{\tiny AU} region with nano-grains regenerated through the photo-fragmentation of large carbonaceous grains and aggregates. This again seems to point away from small arophatic nano-particles as an origin for nano-diamonds. 

It is notable that the 100\,nm radius a-C grains have temperatures $\sim 400-500$\,K at $R = 10-15$\,AU and $\sim 300-400$\,K at $R = 15-30$\,AU, and that they are considerably cooler than the same size nano-diamonds in these regions, that is $\sim 600-2400$\,K and $\sim 500-2100$\,K, respectively. In HR\,4049 these same 100\,nm a-C grain and nano-diamond temperature domains are shifted outwards by about an order of magnitude in radial distance. It is therefore likely that the nano-diamonds could form in situ by the irradiation of these carbon grains. Thus. a better tracer of possible nano-diamond formation sites might then actually be the $\approx 1000$\,K continuum emission arising from the large amorphous carbonaceous grains \citep[e.g.][]{Acke:2013cn,2014ApJ...780...41M}. Nevertheless, and under the \cite{2009ApJ...693..610G} formation scenario, nano-diamonds can only form if the x-ray irradiation conditions are sufficient. Thus, nano-diamonds will only form if all of the conditions are right, principal among them being the x-ray irradiation. The detection of nano-diamonds in the HR\,4049 system, with its apparent lack of x-ray emission, is then something of a conundrum, which might perhaps explain the different form of the $3-4\,\mu$m nano-diamond spectrum in this object. Conversely, the presence of nano-diamonds in this system might indicate the presence of as yet undetected x-ray emission. Once formed in the inner disc regions, the nano-diamonds will be subject to destructive processing and sublimation described in this work and only those formed further out, where there temperatures are lower than $\sim 2000$\,K, the graphitisation threshold, can they survive indefinitely. 

Up to this point it has ben assumed that the nano-diamonds are formed in optically thin regions but we know that the discs themselves are optically thick at UV-IR wavelengths \citep[e.g.][]{2014ApJ...780...41M}. 
Nano-diamond formation within the dense matter of these discs ($n_{\rm H} 10^8-10^{10}$\,cm$^{-3}$) will be inhibited by the extinction of the irradiating x-ray photons, for example if $A_{\rm V} \simeq 0.3$\,mag \citep{Acke:2013cn} then the equivalent x-ray extinction, $A_{\rm x}$, must be of the order of $2-3$\,mag leading to a consequent reduction in the UV and x-ray fluxes by about an order of magnitude, perhaps by enough to inhibit carbon grain transformation to diamond. Thus, the most likely diamond formation sites would seem to be the optically thin regions at disc edges and surfaces. 

Clearly, while the pre-solar meteoritic nano-diamonds ($a = 1-2$\,nm), assumed to be of SN-associated origin, are probably not related to the nano-diamonds observed and/or formed in circumstellar discs ($a \simeq 10-100$\,nm), it nevertheless remains likely that there is some commonality to their formation, that is the necessary x-ray irradiation of hot dust. In this case, and because transformation to diamond must be less than complete, there ought to be a large reservoir of carbon dust with the same isotopic composition. However, this is unlikely to have been preserved because it will almost certainly have been completely re-processed in the ISM \citep[e.g.][]{2011A&A...530A..44J,2014A&A...570A..32B}.

\section{Conclusions}
\label{sect_conclusions}

If nano-diamonds are of extra-solar origin and widespread throughout circumstellar and interstellar regions, then they will be subject to thermal processing and size-sorting in intense radiation field environments, where they ought also to be present.  However, they have to date only been unequivocally detected in less than a handful of sources (e.g. HR\,4049, Elias\,1, and HD\,97048), which argues against their ubiquity, unless it takes special (extreme) environments such as these to make them shine. 

We have here used the recently-derived complex indices of refraction for nano-diamonds as a function of size to predict the nano-diamond temperature profiles in the three sources where they have been detected. Their cooling and hence their temperatures close to these hot stars strongly depend upon the degree of surface hydrogenation, through emission from CH and CH$_2$ bands in the $3-4\,\mu$m and $6-7\,\mu$m wavelength regions and CC band emission at longer wavelengths ($\lambda \simeq 6-9\,\mu$m). 

The critical processes for nano-diamond thermal processing, and their corresponding temperatures, are: 
\begin{itemize}
\item dehydrogenation: loss of H atoms ($T_{\rm nd} \gtrsim 1200-1500$\,K)
\item graphitisation: re-construction to sp$^2$  ($T_{\rm nd} \gtrsim 2000-2100$\,K)
\item sublimation: direct loss of C atoms to the gas ($T_{\rm nd} > 2350$\,K)
\end{itemize}

We find that, in general, the larger nano-diamonds are the hottest and therefore the most susceptible to dehydrogenation, graphitisation, and sublimation close to hot stars. However, at the likely densities of these circumstellar regions, re-hydrogenation will counteract the effects of dehydrogenation. In such regions they should therefore always reveal their presence via surface CH$_n$ emission bands at $3-4\,\mu$m wavelengths. Nevertheless, the results derived here appear to indicate nano-diamond lifetimes against sublimation are of the order of only days for temperatures greater than 1500\,K and that the dynamical timescales, due to radiation pressure-driven drift velocities, are of the order of tens to hundreds of years, that is they are too long to remove them from harm's way. If, however, nano-diamonds can form in the distant disc regions, where the dust is generally cooler, then it is possible that all sizes of nano-diamonds could survive and emit there. 

In the inner, highly excited regions of proto-planetary discs, and perhaps also in some regions of the interstellar medium, e.g. photon-dominated regions (PDRs) with intense and hard ISRFs, it is the smallest nano-diamonds that will likely be the most stable against dehydrogenation, surface and bulk re-construction (graphitisation), and sublimation. In these regions it is therefore most likely a case of the survival of the coolest and smallest nano-diamonds. This may perhaps provide an explanation as to why no large pre-solar nano-diamonds ($a \gg 10$\,nm) have been extracted from primitive meteorites.

Large nano-diamonds ($a \simeq 100$\,nm) are apparently unlikely to survive in highly excited regions because they exhibit low emissivities, and will undergo significant heating leading to graphitisation and fast destruction by sublimation. It is thus not surprising that no large nano-diamonds have yet been detected amongst the extra- and pre-solar grains.  

The results of the modelling presented here appear to highlight an apparently very delicate balance of the processes that regulate nano-diamond survival in energetic regions. If we are to better understand the nature and evolution of nano-diamonds in circumstellar discs it is evident that more precise experimental and theoretical determinations of the nano-diamond optical and thermal properties, and their dynamical behaviour, are urgently required. 
Finally, the results and speculations presented here will need to be explored and tested with fully time-dependent, 3-D radiative transfer models of the gas, dust, and nano-diamonds in proto-planetary discs. 

\begin{acknowledgements}
The author is grateful to the anonymous referee for raising several critical issues that helped to significantly extend and improve the manuscript. 
The author also wishes to thank Nathalie Ysard, Emilie Habart, Emmanuel Dartois and 
numerous other colleagues for interesting discussions relating to nano-diamonds.  \\ \\ 
This work is dedicated to \\ \\ 
Keith, my long-suffering brother of more than 63 years. Taken from us too soon he will remain forever in our hearts and minds. \\ Keith Edward Jones  ( 25$^{th}$ February 1957 $-$  29$^{th}$ October 2020 ) \\ \\
and to \\ \\ 
Dee, a lifelong family friend. \\
Deidre A. Crook (20$^{th}$ January 1933 -- 25$^{th}$ June 2022)

\end{acknowledgements}


\bibliographystyle{bibtex/aa} 
\bibliography{../Ant_bibliography} 

\end{document}